\documentclass[3p,preprint,lefttitle]{elsarticle}

\usepackage{xcolor}
\usepackage[bookmarksnumbered=true,colorlinks=true,citecolor=green!50!black,linkcolor=red!80!black]{hyperref}

\usepackage{amsmath,amssymb,amsthm}

\usepackage[mathlines]{lineno}

\newcommand*\patchAmsMathEnvironmentForLineno[1]{%
	\expandafter\let\csname old#1\expandafter\endcsname\csname #1\endcsname
	\expandafter\let\csname oldend#1\expandafter\endcsname\csname end#1\endcsname
	\renewenvironment{#1}%
	{\linenomath\csname old#1\endcsname}%
	{\csname oldend#1\endcsname\endlinenomath}}%
\newcommand*\patchBothAmsMathEnvironmentsForLineno[1]{%
	\patchAmsMathEnvironmentForLineno{#1}%
	\patchAmsMathEnvironmentForLineno{#1*}}%
\patchBothAmsMathEnvironmentsForLineno{equation}%
\patchBothAmsMathEnvironmentsForLineno{align}%
\patchBothAmsMathEnvironmentsForLineno{flalign}%
\patchBothAmsMathEnvironmentsForLineno{alignat}%
\patchBothAmsMathEnvironmentsForLineno{gather}%
\patchBothAmsMathEnvironmentsForLineno{multline}%

\usepackage[todonotes={textsize=tiny}]{changes}
\definechangesauthor[name={Zhicheng HU}, color=red]{zc}

\setdeletedmarkup{{\color{blue} \sout{#1}}}

\usepackage{float}

\usepackage{subfigure}
\usepackage{cleveref}
\crefname{equation}{}{}
\crefname{algorithm}{Algorithm}{Algorithm}
\crefname{table}{Table}{Table}
\crefname{figure}{\figurename}{\figurename}
\crefname{appendix}{}{}
\crefname{section}{}{}

\usepackage{booktabs}  

\graphicspath{{images}}

\newcommand\pd[2]{\dfrac{\partial {#1}}{\partial {#2}}}
\newcommand\pdtw[2]{\dfrac{\partial^2 {#1}}{\partial {#2}^2}}

\newcommand\dd{\,\mathrm{d}}

\newcommand\mL{\mathcal{L}}

\begin{document}
	\hypersetup{citecolor=green!50!black,linkcolor=red!80!black}
	\begin{frontmatter}
		
		\title{Dual-phase-lag heat conduction analysis of a three-dimensional finite medium heated by a moving laser beam with circular or annular cross-section}
		
		\author[a,b]{Kaiyuan Chen}
		\ead{chenkaiyuan2000@163.com}
		
		\author[a,b]{Longkun Fan}
		\ead{flk2870388597@163.com}
		
		\author[a,b]{Zhicheng Hu\corref{cor1}}
		\ead{huzhicheng@nuaa.edu.cn}
		
		\author[a]{Yixin Xu}
		\ead{2231734756@qq.com}

		\cortext[cor1]{Corresponding author.}
		\affiliation[a]{organization={School of Mathematics, Nanjing University
				of Aeronautics and Astronautics},
			city={Nanjing},
			postcode={211106},
			country={China}}
		
		\affiliation[b]{organization={Key Laboratory of Mathematical
				Modelling and High Performance Computing of Air Vehicles
				(NUAA), MIIT},
			city={Nanjing},
			postcode={211106},
			country={China}}
		
		\begin{abstract}
			We analyze the non-Fourier dual-phase-lag heat conduction process in a three-dimensional medium heated by a moving circular or annular laser beam, which is modeled by a set of point heat sources in the cross-section. In order to solve the model, Green's function approach is first used to obtain an analytical solution for the temperature distribution over the medium subjected to a single point heat source. Then the temperature distribution on the medium subjected to the laser beam can be obtained by the superposition method. According to this solution, the dependence between the heat conduction process and the cross-section of the heat source is investigated. Based on the comparison of the temperature distribution of the medium under Fourier's law and non-Fourier's law, the effect of the phase lag parameter is revealed. In addition, the effects of laser spot size and laser moving speed on the temperature distribution are also analyzed. The discovered properties provide theoretical support for the application of moving laser heat sources in various fields under the dual-phase-lag model.	
		\end{abstract}
		
	\end{frontmatter}
	
\section{Introduction}
\label{sec:intro}
The phenomenon of heat transfer is common in life, and its essence is the thermal movement of molecules, which usually transfers heat from high-temperature molecules to low-temperature molecules. The research of heat transfer processes can help us to better utilize and control the heat, and thus improve the energy utilization efficiency and the quality of life.

In the early 19th century, the French physicist Fourier \cite{fourier1888theorie} proposed Fourier's law, which becomes the classical theory for describing heat transfer processes in modern thermodynamics. Fourier's law, which describes the heat conduction process well in the classical case, assumes that heat conduction is a transient process and heat travels at an infinite velocity. In extreme temperature conditions or at microscopic scales, however, the heat transfer velocity is finite such that the assumption of Fourier propagation cannot be satisfied. Due to the limitations of the classical Fourier's law, some researchers have considered the effect of time lag to modify this fundamental law as follows. 
By introducing a time lag parameter $\tau_{q}$ for the heat flux, Cattaneo \cite{cattaneo1958form} and Vernotte \cite{vernotte1958paradoxes} established so-called CV model to describe the non-constant heat transfer process accurately.
Although $\tau_{q}$ is introduced in the CV model to consider the assumption of finite heat wave propagation, the CV model does not consider the effect of microstructure interactions. Later, Tzou \cite{tzou} further corrected it by introducing a time lag parameter $\tau_{T}$ for the temperature gradient. The resulting final form is referred to as the dual-phase-lag (DPL) model. This model takes into account not only the time lag effect but also the effect of nonlinear heat conduction, and can describe the heat conduction process under the non-Fourier's law more accurately. The CV model and DPL model drive the development of non-Fourier's law and provide reliable theoretical support for the study of heat transfer problems in special situations.
For examples, Ghazanfarian et al. \cite{ghazanfarian2012investigation} applied the DPL model to the modeling of nanoscale thermal transport, which is a key step from an unrealistic transistor condition to the thermal design of real transistor devices in microelectronics. Rukolaine \cite{rukolaine2014unphysical} investigated the nonphysical phenomena of the DPL model in the form of Jeffreys-type equations based on a first-order approximation of the dual-phase-lag constitutive relation. Ghasemi et al. \cite{ghasemi2022dual} provided direct multivariate analytical solutions based on non-Fourier laws for cylinders with any boundary conditions and functional gradient materials.

The problem of moving heat sources, in which the high energy density property of the heat source allows the laser beam to transfer energy in a very short period of time, is an important heat transfer phenomenon and has many applications \cite{hahn2012heat,panas2014moving,zhang2007nano}. In industrial manufacturing, the heat sources are used for fine heating during processing such as surface hardening, cladding, cutting and welding. In biomedical field, moving laser heat sources can be used for corneal cutting and thermal treatment of tumor cells, targeted cutting and repair of human tissues, thermal imaging and spot removal, and other therapeutic processes. Depending on the practical application, the heat source can be modeled as a point, line, surface or volumetric shaped heat source that can be changed in shape and motion to meet the needs of different fields. 
There are many researches focusing on the thermal responses caused by different heat sources. 
Wang et al. \cite{wang2019scaling} investigated the relevant eigenvalues of moving point heat sources on semi-infinite solids based on the intrinsic nature of physics.
Kukla \cite{2008} studied the temperature distribution on the surface of a three-dimensional (3D) medium heated by a curved moving laser heat source.
Zubair and Chaudhry \cite{zubair1996temperature} discussed a closed-form model for the computation of temperature distribution in an infinitely extended isotropic body with a time-dependent moving-line-heat source.
Laraqi \cite{laraqi2003exact} provided an exact explicitly analytical solution of the steady-state temperature in a half space subjected to a moving circular heat source.
Araya and Gutierrez \cite{araya2006analytical} presented an analytical solution of the transient temperature distribution in a finite solid when it was heated by a moving laser beam.

To more accurately describe the heat transfer behavior in special cases such as microscale, a large amount of work involving moving heat sources and the DPL heat conduction model has been carried out by researchers. 
Ma et al. \cite{2018} investigated the temperature effect of a moving delta point heat source on a two-dimensional (2D) medium under the dual-phase-lag model. 
Yang et al. \cite{yang2023transient} used the DPL heat conduction theory to analyze the transient thermal processes in cracked strip materials under ultrafast Gaussian laser beam radiation. 
Shen and Zhang \cite{2022} analyzed the thermal behavior of a two-dimensional plate heated by an annular laser pulse heat source based on the DPL model. 
Considering the trajectory of the heat source, Chen and Hu \cite{chen2023} analyzed the heat transfer of a plate affected by the heat source of elliptical motion based on the DPL model.
When the heat source is fixed and the relative movement of the medium is considered, Lee et al. \cite{lee2016numerical} applied the DPL model to study the transient heat transfer of a moving finite medium under the action of a time-varying laser heat source.
In addition, since the dual-phase-lag model can more accurately describe the heat transfer process on biological tissues, the moving heat source problems have also received considerable attention in this field. Ma et al. \cite{ma2021theoretical} studied the thermal response on the skin surface using a repetitive pulse laser heat source based on the DPL bio-heat transfer model. Partovi et al. \cite{partovi2023analytical} analyzed the transient temperature distribution in a three-dimensional living tissue affected by a moving multi-point laser beam with a square cross-section.

Obviously, it is required to solve the model to obtain the temperature distribution. For the majority of cases such as special initial value conditions and boundary conditions, analytical methods are popular approaches. For instance, 
Frankel et al. \cite{Frankel} used the Green's function method to derive an analytical solution for the temperature distribution of a one-dimensional medium based on the CV model. 
In analogy to the above derivation for a one-dimensional medium, Ma et al. \cite{2018} used the Green's function method to find the temperature distribution on the surface of a two-dimensional medium subjected to a moving point heat source under the DPL model.
Kabiri and Talaee \cite{kabiri2021thermal} used the eigenvalue method to obtain a closed form solution for the temperature field of a moving heat source in cancer thermal therapy.
In addition, numerical methods are also popular, especially for solving complex problems. For the classical Fourier cases, various numerical methods are applicable, while for the DPL model, the finite difference method is still used to derive numerical solutions \cite{torabi2011analytical,sharma2021numerical,ramos2023mathematical,ozicsik2017finite}.

In the previous literature, most of the moving heat source problems incorporating DPL model have been solved for point heat source to obtain the analytical solution of the model. The laser beam is to a certain advantage for micro-scale studies such as laser processing of metallic materials and thermal therapy in biomedicine.
There are important applications for laser beam heat sources with special cross-sections, e.g., Shen and Zhang \cite{2022} studied a two-dimensional medium heated by a moving annular laser based on the DPL model. Therein, Green's function and superposition methods are used to obtain the temperature distribution on the two-dimensional medium. 
Compared to the 2D model, a 3D model is obviously more accurate in simulating the medium heated by moving laser heat sources, since it realistically considers heat transfer in the longitudinal direction. In this paper, we conduct a dual-phase-lag heat transfer analysis of a three-dimensional medium heated by a moving laser beam with circular or annular cross-section. We not only analyze the temperature on the surface of the three-dimensional medium, but also focus on the temperature variation inside the medium. 
The laser beam is simulated by the superposition of point heat sources over the cross-section. Then the heat transfer process is analyzed by the temperature distribution over the medium.

The present paper is built up as follows. The control equation of the problem and the specific manifestation of the heat source are shown in Section \cref{sec:model}, while the Green's function method and finite difference method used to solve this equation are presented in Section \cref{sec:sol}. The comparison of the analytical and numerical solutions, the comparison of Fourier and non-Fourier models, and the discussion of several parameters affecting the temperature distribution of the medium are given in Section \cref{sec:result}. In the final Section \cref{sec:conclusion}, we summarize the main results for the 3D medium subjected to a moving laser beam.

\section{Dual-phase-lag model with a moving laser beam}
\label{sec:model}

According to the generalized DPL model proposed by Tzou \cite{tzou}, the constitutive relation can be written as
\begin{equation}
	\label{eq:nonFourier-law}
	\boldsymbol{q}(x,y,z,t+\tau_{q}) = -k \nabla T(x,y,z,t+\tau_{T}),
\end{equation}
where $\boldsymbol{q}(x,y,z,t)$ is heat flux vector, $T(x,y,z,t)$ is temperature, $k$ is the thermal conductivity, $\tau_{q}$ is the heat flux phase lag and $\tau_{T}$ is the temperature gradient phase lag. Performing the first-order Taylor expansion on \cref{eq:nonFourier-law} yields
\begin{equation}
	\label{eq:1st-dpl-law}
	\boldsymbol{q}(x,y,z,t) + \tau_{q} \frac{\partial \boldsymbol{q}(x,y,z,t)}{\partial t} = - k \left( \nabla T (x,y,z,t) + \tau_{T} \frac{\partial \nabla T(x,y,z,t)}{\partial t} \right).
\end{equation}
Then combining the above with the energy equation, we can get the heat conduction equation of DPL model as
\begin{equation}
	\label{eq:dpl-heat-conduction}
	\nabla \cdot \nabla T + \tau_{T} \pd{}{t}(\nabla \cdot \nabla T) + \frac{1}{k} \left( Q + \tau_{q} \pd{Q}{t} \right) = \frac{1}{\alpha} \left( \pd{T}{t} + \tau_{q} \pdtw{T}{t} \right),
\end{equation}
where $\alpha= k/(\rho C)$ is the thermal diffusivity of the medium, $\rho$ is the material density, and $C$ is the heat capacity.

\begin{figure}[!htbp] 
	\centering
	\includegraphics[width=0.5\textwidth]{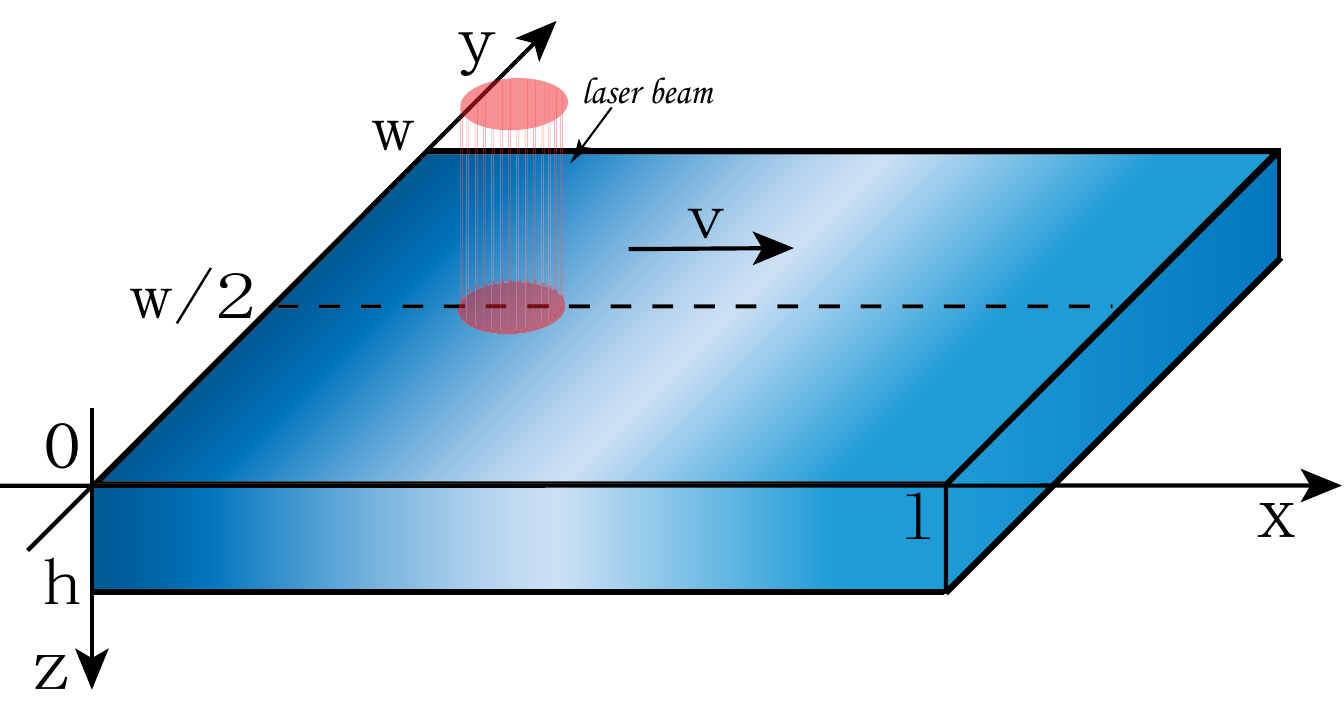}
	\caption{Schematic diagram of the three-dimensional medium with a moving laser beam.}
	\label{fig:sketch} 
\end{figure}

Consider a homogeneous and isotropic cuboid medium heated by a moving heat source, as shown in \cref{fig:sketch}. The heat source can be modeled as a laser beam with a circular or annular cross-section, then the heat transfer process can be simplified as a laser beam heating a 3D medium, where the length, width, and height of the medium are $l$, $w$, and $h$, respectively.

Since our laser beam is simulated by the superposition of point heat sources over the cross-section, it is necessary to solve the temperature distribution over the medium heated by a single point heat source, which e.g. \cite{2022,ma2019thermal,winczek2010analytical}, can be expressed as 
\begin{equation}
	\label{eq:heat-source}
	Q(x,y,z,t) = \mu_{a}(1-R_{a})I_{0}\exp\left(-\mu_{a}z\right) \exp\left(-2\frac{(x-\frac{l}{5}-vt)^2+(y-\frac{w}{2})^2}{r^2}\right),
\end{equation}
where $\mu_{a}$ is energy absorption rate across the depth of the medium, $R_{a}$ represents the light reflectance ratio, $I_{0}$ is the power density of the input laser energy, $v$ represents the moving speed of the heat source and $\left(l/5,w/2,0\right)$ is the initial position of the center of the laser beam. The selected point heat source moves along the positive x-direction with Gaussian distribution on top surface and exponential decay in the longitudinal direction.

So the control equation for this problem is as follows
\begin{equation}
	\label{eq:dpl-heat-conduction-3D}
	\left(\pdtw{T}{x} + \pdtw{T}{y}+ \pdtw{T}{z} \right) + \tau_{T} \left( \pd{^{3} T}{t\partial x^{2}} + \pd{^{3}T}{t\partial y^{2}} + \pd{^{3}T}{t\partial z^{2}} \right) + \frac{1}{k} \left( Q + \tau_{q} \pd{Q}{t} \right) = \frac{1}{\alpha} \left( \pd{T}{t} + \tau_{q} \pdtw{T}{t} \right).
\end{equation}
We suppose that the environmental temperature surrounding the medium is $T_{0}$, and the top surface is set to be adiabatic. Accordingly, the boundary conditions throughout the time would be expressed as    
\begin{equation}
	\label{eq:boundary-condition}
	T(0,y,z,t) = T(l, y,z,t) = T(x,0,z,t) = T(x,w,z,t) = T(x,y,h,t) = T_{0}, \qquad
	\pd{T(x,y,0,t)}{z} = 0.
\end{equation}
It is also assumed that the temperature of the medium under the initial moment is normal temperature and the derivative of the temperature with respect to time is zero. Therefore, the initial time can be given as
\begin{equation}
	\label{eq:initial-condition}
	T(x,y,z,0) = T_{0}, \qquad 
	\pd{T(x,y,z,0)}{t} = 0.
\end{equation}

By introducing the following dimensionless parameters
$$\theta = \frac{T-T_{0}}{T_{0}},\quad \tau = \frac{t}{\tau_q},\quad K_{\tau} = \frac{\tau_{T}}{\tau_{q}},\quad \xi_{1} = \frac{x}{\sqrt{\alpha \tau_{q}}},\quad \xi_{2} = \frac{y}{\sqrt{\alpha \tau_{q}}},\quad \xi_{3} = \frac{z}{\sqrt{\alpha \tau_{q}}},\quad V = \frac{v}{\sqrt{\alpha/\tau_{q}}},$$
$$ L = \frac{l}{\sqrt{\alpha \tau_{q}}},\quad W = \frac{w}{\sqrt{\alpha \tau_{q}}},\quad  H = \frac{h}{\sqrt{\alpha \tau_{q}}},\quad R = \frac{r}{\sqrt{\alpha \tau_{q}}},\quad R_{1} = \frac{r_{1}}{\sqrt{\alpha \tau_{q}}},\quad R_{2} = \frac{r_{2}}{\sqrt{\alpha \tau_{q}}},$$
we get the dimensionless control equation and the initial-boundary conditions are as follows
\begin{equation}
	\label{eq:dpl-heat-conduction-dim}
	\left(\pdtw{\theta}{\xi_{1}} + \pdtw{\theta}{\xi_{2}}+ \pdtw{\theta}{\xi_{3}} \right) + K_{\tau} \left( \pd{^{3} \theta}{\tau \partial \xi_{1}^{2}} + \pd{^{3}\theta}{\tau \partial \xi_{2}^{2}} + \pd{^{3}\theta}{\tau \partial \xi_{3}^{2}} \right) + q = \pd{\theta}{\tau} + \pdtw{\theta}{\tau} , 
\end{equation}
\begin{equation}
	\label{eq:initial-condition-dim}
	\theta(\xi_{1},\xi_{2},\xi_{3},0) = 0, \qquad 
	\pd{T(\xi_{1},\xi_{2},\xi_{3},0)}{\tau} = 0,
\end{equation}
\begin{equation}
	\label{eq:boundary-condition-dim}
	\theta(0,\xi_{2},\xi_{3},\tau) = \theta(L,\xi_{2},\xi_{3},\tau) = \theta(\xi_{1},0,\xi_{3},\tau) = \theta(\xi_{1},W,\xi_{3},\tau) = \theta(\xi_{1},\xi_{2},H,\tau) = 0, \qquad
	\pd{\theta(\xi_{1},\xi_{2},0,\tau)}{\xi_{3}} = 0,
\end{equation}
where the heat from a point heat source can be expressed as
\begin{equation}
	\label{eq:q-dim}
	q\left(\xi_{1},\xi_{2},\xi_{3},\tau \right)=\frac{\alpha \tau_{q}}{kT_{0}} \mu_{a}(1-R_{a})I_{0}\left( 1+\frac{4V\left( \xi_{1}-\frac{L}{5}-V\tau \right)}{R^2} \right)\exp\left(-\mu_{a} \sqrt{\alpha \tau_{q}} \xi_{3}\right) \exp\left(-2\frac{(\xi_{1}-\frac{L}{5}-V\tau)^2+(\xi_{2}-\frac{W}{2})^2}{R^2}\right) .
\end{equation}

\section{Formulation of the solution}
\label{sec:sol}
In this section, we solve for the temperature of a 3D medium heated by a moving laser beam under the DPL heat conduction model. The focus is on the analytical solution of the temperature distribution, which can be solved first by Green's function method for the temperature under the action of a single point heat source. Then the analytical solution for the temperature of the 3D medium under the action of the laser beam is obtained by integrating over a circular or annular region according to the superposition method. Finally, the finite difference method is used for discretization to prepare for the subsequent comparison of the analytical and numerical solutions.

\subsection{Series representation of analytical solution}
Green's function method for the CV model in one-dimensional media \cite{Frankel} is extended to the DPL model in three-dimensional finite media to obtain general solutions for related heat conduction problems.
Denote a modified operator $\mL$ as 
\begin{equation}
	\label{eq:operator-L}
	\mL =  \pd{}{\tau} + \pdtw{}{\tau}  - \left(\pdtw{}{\xi_{1}} + \pdtw{}{\xi_{2}} + \pdtw{}{\xi_{3}} \right) - K_{\tau} \left( \pd{^{3}}{\tau \partial \xi_{1}^{2}} + \pd{^{3}}{\tau \partial \xi_{2}^{2}} + \pd{^{3}}{\tau \partial \xi_{3}^{2}} \right).
\end{equation}
It follows that the DPL heat conduction equation \cref{eq:dpl-heat-conduction-dim} can be rewritten as
\begin{equation}
	\label{eq:dpl-heat-eqnew}
	\mL \theta \left(\xi_{1},\xi_{2},\xi_{3},\tau \right) = q\left(\xi_{1},\xi_{2},\xi_{3},\tau \right).
\end{equation}
The operator $\mL'$ of the cause variables $\xi_{1}',\xi_{2}',\xi_{3}'$ and $\tau'$ in the Green's function can be expressed as
\begin{equation}
	\label{eq:operator-Lxi}
	\mL' =  \pd{}{\tau'} + \pdtw{}{\tau'}  - \left(\pdtw{}{\xi_{1}'} + \pdtw{}{\xi_{2}'} + \pdtw{}{\xi_{3}'} \right) - K_{\tau} \left( \pd{^{3}}{\tau' \partial {\xi_{1}'}^{2}} + \pd{^{3}}{\tau' \partial {\xi_{2}'}^{2}} + \pd{^{3}}{\tau' \partial {\xi_{3}'}^{2}} \right).
\end{equation}

Now let us consider the integration given by
\begin{equation}
	\label{eq:integral-def}
	I = \lim_{\varepsilon\to 0} \int_{0}^{t+\varepsilon} \int_{0}^{L} \int_{0}^{W} \int_{0}^{H} G(\xi_{1},\xi_{2},\xi_{3},\tau|\xi_{1}',\xi_{2}',\xi_{3}',\tau') \mL'\theta(\xi_{1}',\xi_{2}',\xi_{3}',\tau') \dd \xi_{3}' \dd \xi_{2}' \dd \xi_{1}' \dd \tau',
\end{equation}
where $G(\xi_{1},\xi_{2},\xi_{3},\tau|\xi_{1}',\xi_{2}',\xi_{3}',\tau')$ is the appropriate Green's function representing  the ``effect/cause" relationship, and the $\varepsilon$ is a small value which is introduced to invoke the above relationship in the following.

According to the integration by parts, we could rewrite it as
\begin{equation}
	\label{eq:integral-sum}
	I = \lim_{\varepsilon \to 0} \left(I_{1} + I_{2} + I_{3} + I_{4} + I_{5}\right),
\end{equation}
where
\begin{align}
	\label{eq:integral-I1}
	& I_{1} = \int_{0}^{L} \int_{0}^{W} \int_{0}^{H} \left.\left( G\theta +  G \pd{\theta}{\tau'} - \pd{G}{\tau'} \theta - K_{\tau} G\left( \pdtw{\theta}{\xi_{1}'} +\pdtw{\theta}{\xi_{2}'} +\pdtw{\theta}{\xi_{3}'} \right) \right)\right|_{0}^{\tau+\varepsilon}\dd \xi_{3}' \dd \xi_{2}' \dd \xi_{1}',\\
	& I_{2} = \int_{0}^{\tau+\varepsilon} \int_{0}^{W} \int_{0}^{H} \left.\left( -G \pd{\theta}{\xi_{1}'} + \pd{G}{\xi_{1}'} \theta + K_{\tau} \pd{G}{\tau'} \pd{\theta}{\xi_{1}'} - K_{\tau} \pd{^{2}G}{\xi_{1}' \partial \tau'} \theta \right)\right|_{0}^{L} \dd \xi_{3}' \dd \xi_{2}' \dd \tau', \label{eq:integral-I2}\\
	& I_{3} = \int_{0}^{\tau+\varepsilon} \int_{0}^{L} \int_{0}^{H} \left.\left( -G \pd{\theta}{\xi_{2}'} + \pd{G}{\xi_{2}'} \theta + K_{\tau} \pd{G}{\tau'} \pd{\theta}{\xi_{2}'} - K_{\tau} \pd{^{2}G}{\xi_{2}' \partial \tau'} \theta \right)\right|_{0}^{W} \dd \xi_{3}' \dd \xi_{1}' \dd \tau', \label{eq:integral-I3}\\
	& I_{4} = \int_{0}^{\tau+\varepsilon} \int_{0}^{L} \int_{0}^{W} \left.\left( -G \pd{\theta}{\xi_{3}'} + \pd{G}{\xi_{3}'} \theta + K_{\tau} \pd{G}{\tau'} \pd{\theta}{\xi_{3}'} - K_{\tau} \pd{^{2}G}{\xi_{3}' \partial \tau'} \theta \right)\right|_{0}^{H} \dd \xi_{2}' \dd \xi_{1}' \dd \tau', \label{eq:integral-I4}\\
	& I_{5} = \int_{0}^{\tau+\varepsilon}\int_{0}^{L}\int_{0}^{W}\int_{0}^{H}  \mL'^{*} \left(G(\xi_{1},\xi_{2},\xi_{3},\tau|\xi_{1}',\xi_{2}',\xi_{3}',\tau') \theta(\xi_{1}',\xi_{2}',\xi_{3}',\tau')\right) \dd \xi_{3}' \dd \xi_{2}' \dd \xi_{1}' \dd \tau'. \label{eq:integral-I5}
\end{align}

It should be noted that, the chosen Green's function $G(\xi_{1},\xi_{2},\xi_{3},\tau|\xi_{1}',\xi_{2}',\xi_{3}',\tau')$ is supposed to meet the following equation,
\begin{equation}
	\label{eq:green-eq}
	\mL'^{*} G(\xi_{1},\xi_{2},\xi_{3},\tau|\xi_{1}',\xi_{2}',\xi_{3}',\tau') = \delta(\xi_{1}-\xi_{1}') \delta(\xi_{2}-\xi_{2}') \delta(\xi_{3}-\xi_{3}') \delta(\tau-\tau'),
\end{equation}
where the adjoint operator $\mL'^{*}$ of $\mL'$ is represented as
\begin{equation}
	\label{eq:operator-L-adjoint}
	\mL'^{*} = -\pd{}{\tau'} + \pdtw{}{\tau'}  - \left(\pdtw{}{\xi_{1}'} + \pdtw{}{\xi_{2}'} + \pdtw{}{\xi_{3}'} \right) + K_{\tau} \left( \pd{^{3}}{\tau' \partial {\xi_{1}'}^{2}} + \pd{^{3}}{\tau' \partial {\xi_{2}'}^{2}} + \pd{^{3}}{\tau' \partial {\xi_{3}'}^{2}} \right).
\end{equation}
Meanwhile, \cref{eq:green-eq} should satisfy the following uniform boundary conditions and additional conditions,
\begin{equation}
	\label{eq:green-eq-bc}
	\begin{aligned}
		&G(\xi_{1},\xi_{2},\xi_{3},\tau|0,\xi_{2}',\xi_{3}',\tau') = G(\xi_{1},\xi_{2},\xi_{3},\tau|L,\xi_{2}',\xi_{3}',\tau') = G(\xi_{1},\xi_{2},\xi_{3},\tau|\xi_{1}',0,\xi_{3}',\tau') =\\ &G(\xi_{1},\xi_{2},\xi_{3},\tau|\xi_{1}',W,\xi_{3}',\tau') = 
		G(\xi_{1},\xi_{2},\xi_{3},\tau|\xi_{1}',\xi_{2}',H,\tau') =
		\pd {G(\xi_{1},\xi_{2},\xi_{3},\tau|\xi_{1}',\xi_{2}',0,\tau')}{\xi_{3}'}
		=0,
	\end{aligned}
\end{equation}

\begin{equation}
	\label{eq:green-eq-ic}
	G(\xi_{1},\xi_{2},\xi_{3},\tau|\xi_{1}',\xi_{2}',\xi_{3}',\tau') = \pd{G(\xi_{1},\xi_{2},\xi_{3},\tau|\xi_{1}',\xi_{2}',\xi_{3}',\tau')}{\tau'} = 0, \quad \tau<\tau'.
\end{equation}

Since the initial conditions of the temperature, the additional
conditions of the Green's function, and the boundary conditions of them
are all assumed to be zero, it is easy to show that
$I_{1}=I_{2}=I_{3}=I_{4} = 0$. As for the integral $I_{5}$, substituting
\cref{eq:green-eq} into \cref{eq:integral-I5} and using the properties
of the Dirac delta function, we obtain that $I_{5} =
\theta(\xi_{1},\xi_{2},\xi_{3},\tau)$. 
By noting that $\mL' \theta(\xi_{1}',\xi_{2}',\xi_{3}',\tau') = q(\xi_{1}',\xi_{2}',\xi_{3}',\tau')$,
the temperature can ultimately be represented by the relevant Green's function,
\begin{equation}
	\label{eq:integral-form-sol}
	\theta(\xi_{1},\xi_{2},\xi_{3},\tau) = I = \int_{0}^{\tau} \int_{0}^{L} \int_{0}^{W} \int_{0}^{H} G(\xi_{1},\xi_{2},\xi_{3},\tau|\xi_{1}',\xi_{2}',\xi_{3}',\tau') q(\xi_{1}',\xi_{2}',\xi_{3}',\tau') \dd \xi_{3}' \dd \xi_{2}' \dd \xi_{1}' \dd \tau'.
\end{equation}

In order to obtain the expression of temperature, the first step is to obtain the relevant Green's function. The equation and conditions related to this function are \cref{eq:green-eq,eq:green-eq-bc,eq:green-eq-ic}, so the characteristic functions that satisfy the above boundary conditions are
\begin{equation}
	\label{eq:trig-eig-fun}
	\begin{aligned}
		&X_{m}(\xi_{1})=\sin\alpha_{m}\xi_{1}, \quad \alpha_{m}=\frac{m\pi}{L},\\
		&Y_{n}(\xi_{2})=\sin\beta_{n}\xi_{2}, \quad \beta_{n}=\frac{n\pi}{W},\\
		&Z_{s}(\xi_{3})=\cos\gamma_{s}\xi_{3}, \quad \gamma_{s}=\frac{(2s-1)\pi}{2H}.
	\end{aligned}
\end{equation}
Similar to \cite{2018,2022}, it follows that the Green's function of
the current problem can be expanded into a series in terms of these
eigenfunctions, that is,
\begin{equation}
	\label{eq:green-fun-series-def}
	G(\xi_{1},\xi_{2},\xi_{3},\tau|\xi_{1}',\xi_{2}',\xi_{3}',\tau') = \sum_{m=1}^{\infty}\sum_{n=1}^{\infty}\sum_{s=1}^{\infty} \frac{X_{m}(\xi_{1}') Y_{n}(\xi_{2}') Z_{s}(\xi_{3}')}{M_{m}N_{n}S_{s}} \bar{G}_{mns}(\xi_{1},\xi_{2},\xi_{3},\tau|\xi_{1}',\xi_{2}',\xi_{3}',\tau'),
\end{equation}
where $M_{m}=\int_{0}^{L} X_{m}^{2}(\xi_{1}) \dd \xi_{1}$, $N_{n}=\int_{0}^{W} Y_{n}^{2}(\xi_{2}) \dd \xi_{2}$ and $S_{s}=\int_{0}^{H} Z_{s}^{2}(\xi_{3}) \dd \xi_{3}$. From the orthogonality of eigenfunctions, we get
\begin{equation}
	\label{eq:green-fun-series-coe}
	\bar{G}_{mns}(\xi_{1},\xi_{2},\xi_{3},\tau|\xi_{1}',\xi_{2}',\xi_{3}',\tau') =  \int_{0}^{\tau}\int_{0}^{L}\int_{0}^{W}\int_{0}^{H} G(\xi_{1},\xi_{2},\xi_{3},\tau|\xi_{1}',\xi_{2}',\xi_{3}',\tau') X_{m}(\xi_{1}') Y_{n}(\xi_{2}') Z_{s}(\xi_{3}') \dd \xi_{3}' \dd \xi_{2}' \dd \xi_{1}' \dd \tau'.
\end{equation}

Multiplying both sides of \cref{eq:green-eq} by the function
$X_{m}(\xi_{1})Y_{n}(\xi_{2})Z_{s}(\xi_{3})$, and integrating over the domain
$[0,L]\times[0,W]\times[0,H]$, one can obtain
\begin{equation}
	\label{eq:green-Gmn-ode}
	\pdtw{\bar{G}_{mns}}{\tau'} - \left(1 + K_{\tau} \lambda_{mns}^{2} \right) \pd{\bar{G}_{mns}}{\tau'} + \lambda_{mns}^{2} \bar{G}_{mns} = X_{m}(\xi_{1}) Y_{n}(\xi_{2}) Z_{s}(\xi_{3})\delta(\tau-\tau'),
\end{equation}
where $\lambda_{mns}^{2} = \alpha_{m}^{2} + \beta_{n}^{2} + \gamma_{s}^{2}$. 
Considering the additional conditions obtained from the integral transform of
\cref{eq:green-eq-ic}, the solution for the differential equation \cref{eq:green-Gmn-ode} can be expressed as
\begin{equation}
	\label{eq:green-Gmn-sol}
	\bar{G}_{mns}(\xi_{1},\xi_{2},\xi_{3},\tau|\xi_{1}',\xi_{2}',\xi_{3}',\tau') = \frac{X_{m}(\xi_{1}) Y_{n}(\xi_{2}) Z_{s}(\xi_{3})}{\sigma_{2}} \exp(-\sigma_{1}(\tau-\tau')) \sinh(\sigma_{2}(\tau-\tau')),
\end{equation}
in which
\begin{equation}
	\label{eq:green-Gmn-sol-betas}
	\sigma_{1} = \frac{1 + K_{\tau} \lambda_{mns}^{2}}{2}, \quad \sigma_{2} = \frac{\sqrt{\left(1 + K_{\tau} \lambda_{mns}^{2}\right)^{2} - 4 \lambda_{mns}^{2}}}{2}.
\end{equation}
Substituting \cref{eq:green-Gmn-sol} into \cref{eq:green-fun-series-def} yields the Green's function related to temperature distribution as
\begin{equation}
	\label{eq:green-fun-series-sol}
	G(\xi_{1},\xi_{2},\xi_{3},\tau|\xi_{1}',\xi_{2}',\xi_{3}',\tau') = \sum_{m=1}^{\infty}\sum_{n=1}^{\infty}\sum_{s=1}^{\infty} \frac{X_{m}(\xi_{1}') Y_{n}(\xi_{2}') Z_{s}(\xi_{3}')}{M_{m}N_{n}S_{s}} \frac{X_{m}(\xi_{1}) Y_{n}(\xi_{2}) Z_{s}(\xi_{3})}{\sigma_{2}} \exp(-\sigma_{1}(\tau-\tau')) \sinh(\sigma_{2}(\tau-\tau')).  
\end{equation}

Since the boundary conditions are homogeneous and the initial conditions are zero in this problem, substituting
\cref{eq:green-fun-series-sol,eq:q-dim} into
\cref{eq:integral-form-sol}, the solution for \cref{eq:dpl-heat-conduction-dim} can be written as

\begin{equation}
	\label{eq:series-form-sol}
	\theta(\xi_{1},\xi_{2},\xi_{3},\tau) = \sum_{m=1}^{\infty}\sum_{n=1}^{\infty}\sum_{s=1}^{\infty} \frac{\alpha \tau_{q} \mu_{a} (1-R_{a})I_{0}}{M_{m}N_{n}S_{s} \sigma_{2} k T_0} F_{n} F_{s} F_{mns}(\tau) X_{m}(\xi_{1}) Y_{n}(\xi_{2}) Z_{s}(\xi_{3}),  
\end{equation}
where
\begin{align}
	\label{eq:series-form-sol-coe}
	\begin{aligned}
		&F_{n}=\int_{0}^{W} Y_{n}(\xi_{2}') \exp\left(-2\frac{(\xi_{2}'-\frac{W}{2})^2}{R^2}\right) \dd \xi_{2}',\quad 
		F_{s}=\int_{0}^{H} Z_{s}(\xi_{3}') \exp\left(-\mu_{a} \sqrt{\alpha \tau_{q}} \xi_{3}' \right) \dd \xi_{3}',\\
		&F_{mns}(\tau)=\int_{0}^{\tau}\int_{0}^{L} X_{m}(\xi_{1}') \left( 1+\frac{4 V (\xi_{1}'-\frac{L}{5}-V \tau')}{R^2} \right) \exp\left(-2\frac{(\xi_{1}'-\frac{L}{5}-V \tau')}{R^2}\right)  \exp(-\sigma_{1}(\tau-\tau')) \sinh(\sigma_{2}(\tau-\tau')) \dd \xi_{1}' \dd \tau'.
	\end{aligned}
\end{align}

The above formula \cref{eq:series-form-sol} is the temperature distribution on the medium subjected to a point heat source. The superposition method demonstrates that the total temperature distribution can be obtained by combining the results of all factors affecting the temperature field. Since we want to obtain different cross-section shapes of laser beam such as circle or ring, we can integrate the set of point heat sources in certain regions. Thus, the temperature distribution generated by a moving circular or annular laser beam can be given as
\begin{equation}
	\theta(\xi_{1},\xi_{2},\xi_{3},\tau) = \sum_{m=1}^{\infty}\sum_{n=1}^{\infty}\sum_{s=1}^{\infty} \frac{\alpha \tau_{q} \mu_{a} (1-R_{a})I_{0} Z_{s}(\xi_{3})}{M_{m}N_{n}S_{s} \sigma_{2} k T_0} F_{n} F_{s} F_{mns}(\tau) 
	\int_{R_{1}}^{R_{2}}\int_{0}^{2\pi} X_{m}(\xi_{1}-R_{0}\cos\theta) Y_{n}(\xi_{2}-R_{0}\sin\theta) R_{0} \dd\theta \dd R_{0}.
\end{equation}
In practice, taking the computational cost and the accuracy into
account, the infinite series of the temperature
\cref{eq:series-form-sol} would be truncated with three appropriately
selected integers $M, N$ and $S$ so that
\begin{equation}
	\label{eq:series-truncated-sol}
	\theta(\xi_{1},\xi_{2},\xi_{3},\tau) \approx \sum_{m=1}^{M}\sum_{n=1}^{N}\sum_{s=1}^{S} \frac{\alpha \tau_{q} \mu_{a} (1-R_{a})I_{0} Z_{s}(\xi_{3})}{M_{m}N_{n}S_{s} \sigma_{2} k T_0} F_{n} F_{s} F_{mns}(\tau) 
	\int_{R_{1}}^{R_{2}}\int_{0}^{2\pi} X_{m}(\xi_{1}-R_{0}\cos\theta) Y_{n}(\xi_{2}-R_{0}\sin\theta) R_{0} \dd\theta \dd R_{0},
\end{equation}
the integral calculation of the above formula can be seen in \cref{appendix}.

\subsection{Finite difference method}
The discretization can be performed in various ways, such as finite difference, finite volume, finite element \cite{ivanovic2011numerical,resendiz2015two,hu2018numerical,hu2020heat,champagne2020numerical}. In this paper, the implicit finite difference method is used to obtain the numerical solution of the temperature distribution of a 3D medium. First, the DPL model \cref{eq:dpl-heat-conduction-dim} is rewritten as the following set of equations,
\begin{equation}
	\label{eq:dpl-heat-conduction-dim-set}
	\begin{cases}
		\left( u + \pd{u}{\tau} \right) - \left(\pdtw{\theta}{\xi_{1}} + \pdtw{\theta}{\xi_{2}} + \pdtw{\theta}{\xi_{3}} \right) - K_{\tau} \left( \pdtw{u}{\xi_{1}} + \pdtw{u}{\xi_{2}} + \pdtw{u}{\xi_{3}} \right)  =  q, \\
		\hspace{3mm} u=\pd{\theta}{\tau}.
	\end{cases}
\end{equation}

\begin{figure}[htbp]
	\centering
	\includegraphics[width=0.4\textwidth]{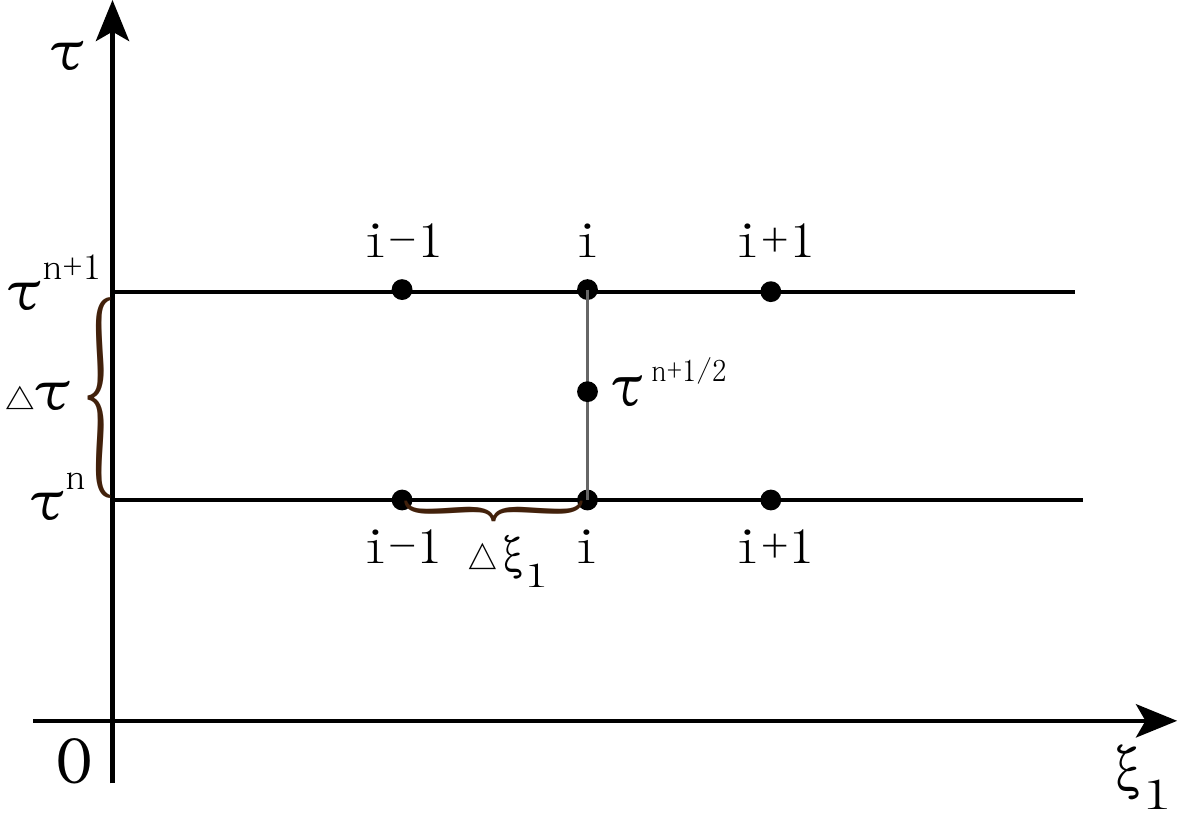}
	\caption{A grid computing model with implicit finite difference in the $\xi_{1}$-direction.}
	\label{fig:differential_space}
\end{figure}

The central difference method and the Crank-Nicolson scheme are used for spatial discretization and temporal discretization respectively, such that the full discretization has second-order accuracy. For a general function $g\left(\xi_{1},\xi_{2},\xi_{3},\tau\right)$, which can denote either $\theta$ or $u$ in \cref{eq:dpl-heat-conduction-dim-set}, the second partial derivative with respect to $\xi_{1}$ and the first partial derivative with respect to $\tau$ are given by
\begin{equation}
	\label{eq:discretization_x}	
	\pdtw{g_{i,j,k}^{n+1/2}}{\xi_{1}} \approx \frac{g_{i+1,j,k}^{n+1/2}-2g_{i,j,k}^{n+1/2}+g_{i-1,j,k}^{n+1/2}}{(\Delta \xi_{1})^2},\quad 
	\pd{g_{i,j,k}^{n+1/2}}{\tau} \approx \frac{g_{i,j,k}^{n+1}-g_{i,j,k}^{n}}{\Delta \tau},
\end{equation}
in which $g_{i,j,k}^{n+1/2} =\frac{g_{i,j,k}^{n}+g_{i,j,k}^{n+1}}{2}$, and 
$g_{i,j,k}^{n} \approx g\left(\xi_{1i},\xi_{2j},\xi_{k3},\tau_{n}\right)$.

\section{Results and discussion}
\label{sec:result}
Following the analytical solution obtained in the previous section, the heat conduction process in the medium under the action of a moving circular or annular laser beam is analyzed in this section. The accuracy of the analytical solution is first verified compared with the numerical solution. Then the effect of the phase lag parameter on the temperature distribution is revealed by comparing three heat conduction models under Fourier's law and non-Fourier's law. Finally, the effects of phase lag ratio, laser spot size, and laser beam moving speed on the surface and internal temperature distribution of the medium are specifically analyzed.
In this work, the physical parameters involved in the medium and laser beam are shown in \cref{tab:default-para,tab:HS_para}, while $\tau_{q}$ takes the value of $1ps$ and has the same order of magnitude as time, and some of these parameters are referenced from the work of \cite{2022}. Taking into account the time and computational cost, the truncated integer 30, used to represent the $M$, $N$ and $S$, is sufficiently accurate for the series \cref{eq:series-truncated-sol}. 
\begin{table}[!htbp]
	\centering
	\caption{Default parameters in the calculation.}
	\label{tab:default-para}
	\begin{tabular}{ll}
		\toprule [2pt]
		Parameters & Value\\
		\midrule
		Absorption coefficient $\mu_{a}$ $(nm^{-1})$ & $0.4$\\ 
		Reflection ration $R_{a}$ & $0.024$\\
		Source energy density $I_0$ $(W\cdot m^{-3})$ & $3.35\times 10^{15}$\\
		Thermal diffusivity $\alpha$ $(m^{2}/s)$ & $1.366\times 10^{-5}$\\
		Thermal conductivity $k$ $(W/(m\cdot K))$ & $49.8$\\
		Environmental temperature $T_0$ $(K)$ & $290$\\
		Moving speed $v$ (m/s) & $5\times 10^{3}$\\
		Distribution parameter $r$ $(nm)$ & $2$\\
		Length $l$ $(nm)$ & $50$\\
		Width $w$ $(nm)$ & $50$\\
		Thickness $h$ $(nm)$ & $20$\\
		\bottomrule [2pt]
	\end{tabular}
\end{table}
\begin{table}[!htbp]
	\centering
	\caption{Laser beam parameters for different cross-section shapes}
	\label{tab:HS_para}
	\begin{tabular}{ccc}
		\toprule
		Parameters & Circular laser beam & Annular laser beam\\
		\midrule  		
		Inner radius $r_1$ $(nm)$ & $0$ & $4$ \\
		Outer radius $r_2$ $(nm)$ & $2$ & $6$ \\
		\bottomrule
	\end{tabular}
\end{table}

\subsection{Comparison of analytical and numerical solutions}
\begin{figure}[!htbp]
	\centering
	\includegraphics[width=0.45\textwidth]{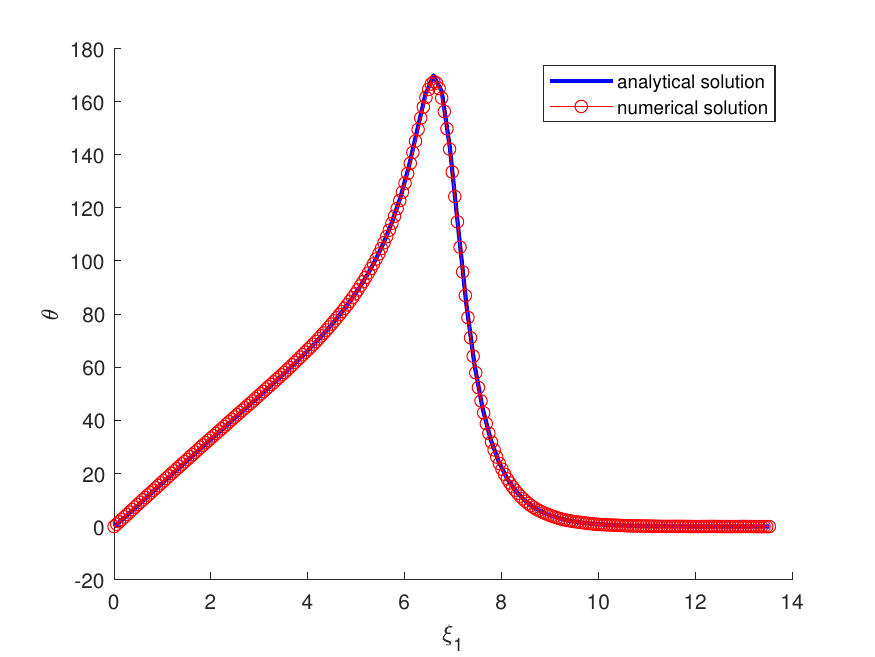}
	\caption{Comparison of analytical and numerical solutions for the top surface temperature distribution on $\xi_{2}=W/2$.}
	\label{fig:com_an}
\end{figure}

In order to verify the above analytical procedure, we used the numerical solution to compare the temperature of the medium subjected by a point heat source under the DPL model. 
In \cref{fig:com_an}, when the heat source moves to the center of the medium, the top surface temperature distribution on $\xi_{2}=W/2$ with the ratio of phase lags $K_{\tau}=1$ is illustrated. 
It is observed in \cref{fig:com_an} that the analytical solution agrees well with the numerical solution, which proves the validity of the Green's function method employed in the present work.

\subsection{Comparison of Fourier and non-Fourier models}

By assuming the temperature gradient phase lag parameter $\tau_{T}$ in the generalized DPL model to be $0$, the model \cref{eq:nonFourier-law} is transformed into a generalized CV model. Setting both the heat flux phase lag parameter $\tau_{q}$ and the temperature gradient phase lag parameter $\tau_{T}$ in the DPL model to be 0, the model degenerates to the classical Fourier model.
Since $\tau_{q}$ appears in the denominator of  series \cref{eq:series-truncated-sol}, which is not representable as a solution to the classical heat conduction model. By reducing the second-order equation \cref{eq:green-Gmn-ode} to the first-order equation, the expression for the temperature distribution based on Fourier's law can be obtained,
\begin{equation}
	\label{eq:series-truncated-sol-fourier}
	\theta(\xi_{1},\xi_{2},\xi_{3},\tau) = \sum_{m=1}^{\infty}\sum_{n=1}^{\infty}\sum_{s=1}^{\infty} \frac{\alpha 	 \mu_{a} (1-R_{a})I_{0} Z_{s}(\xi_{3})}{M_{m}N_{n}S_{s} k T_0} F_{n} F_{s} F_{mns}(\tau) 
	\int_{R_{1}}^{R_{2}}\int_{0}^{2\pi} X_{m}(\xi_{1}-R_{0}\cos\theta) Y_{n}(\xi_{2}-R_{0}\sin\theta) R_{0} \dd\theta \dd R_{0},
\end{equation}
where
\begin{align}
	\label{eq:series-form-sol-fourier-coe}
	\begin{aligned}
		&F_{n}=\int_{0}^{W} Y_{n}(\xi_{2}') \exp\left(-2\frac{(\xi_{2}'-\frac{W}{2})^2}{R^2}\right) \dd \xi_{2}',\quad 
		F_{s}=\int_{0}^{H} Z_{s}(\xi_{3}') \exp\left(-\mu_{a} \sqrt{\alpha} \xi_{3}' \right) \dd \xi_{3}',\\
		&F_{mns}(\tau)=\int_{0}^{\tau}\int_{0}^{L} X_{m}(\xi_{1}') \exp\left(-2\frac{(\xi_{1}'-\frac{L}{5}-V \tau')}{R^2}\right) \exp(-\lambda_{mns}^{2} (\tau-\tau')) \dd \xi_{1}' \dd \tau'.
	\end{aligned}
\end{align}

\begin{figure}[!htbp]
	\centering
	\includegraphics[width=0.45\textwidth]{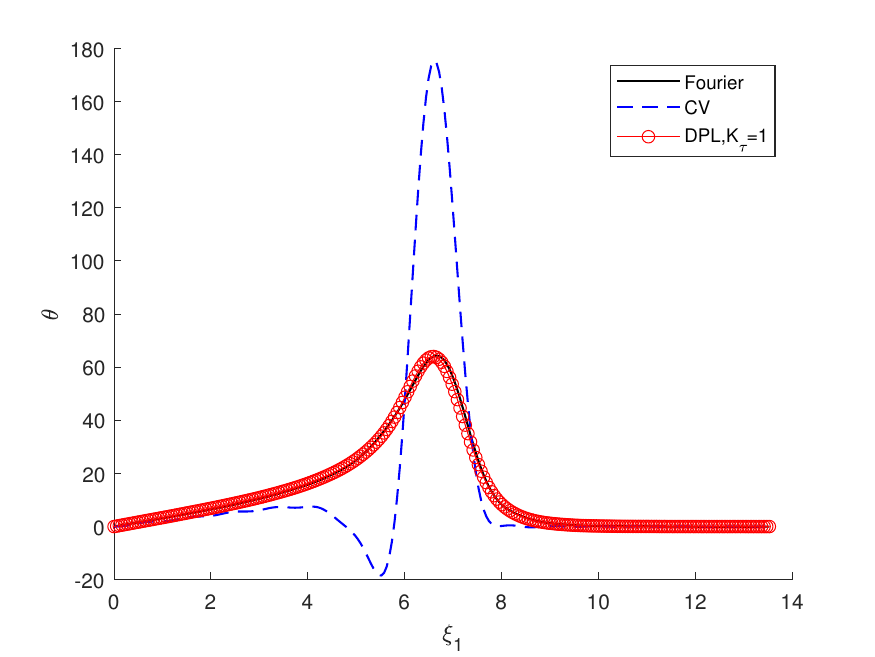}
	\caption{Top surface temperatures on $\xi_{2}=W/2$ with a circular laser beam under different models.}
	\label{fig:com_fd}
\end{figure}

Taking a circular laser beam as an example, it can be found from \cref{fig:com_fd} that when $K_{\tau}=1$, the temperature on $\xi_{2}=W/2$ of the DPL model basically coincides with that of Fourier's law, and this phenomenon can be illustrated in the generalized models. The temperature of the medium under the CV model is obviously different from the Fourier model and the DPL model when $K_{\tau}=0$. In the CV model, the temperature of the heated area is high and the region heated by the laser beam appears unphysical phenomenon due to the effect of the phase lag of the heat flux \cite{liu2007analysis}.

\begin{figure}[!htbp]
	\centering
	\subfigure[The 3D temperature distribution on the surface of the medium.]
	{\includegraphics[width=0.4\textwidth]{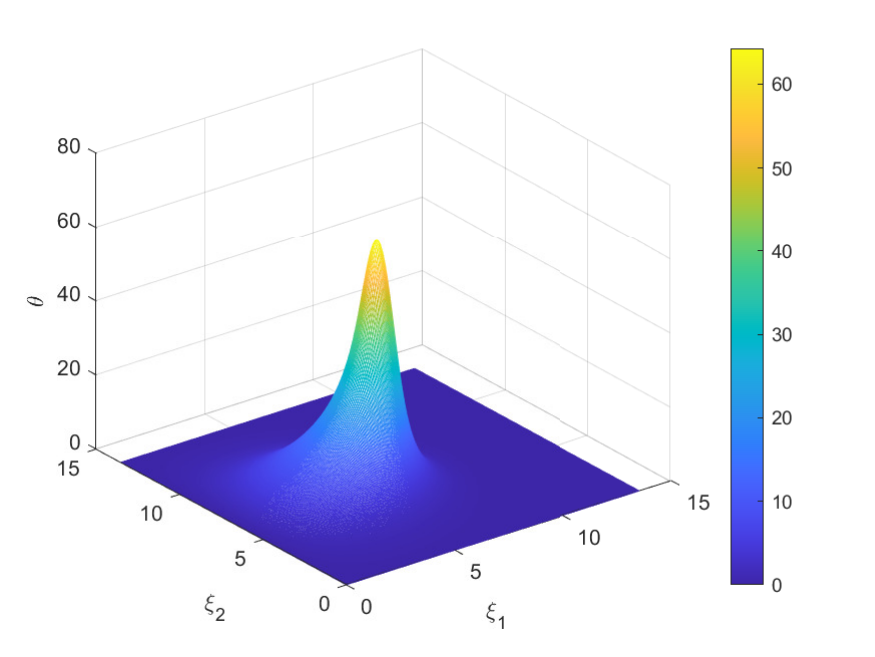}}\\
	\subfigure[The contour of the temperature on the top surface.]
	{\includegraphics[width=0.35\textwidth]{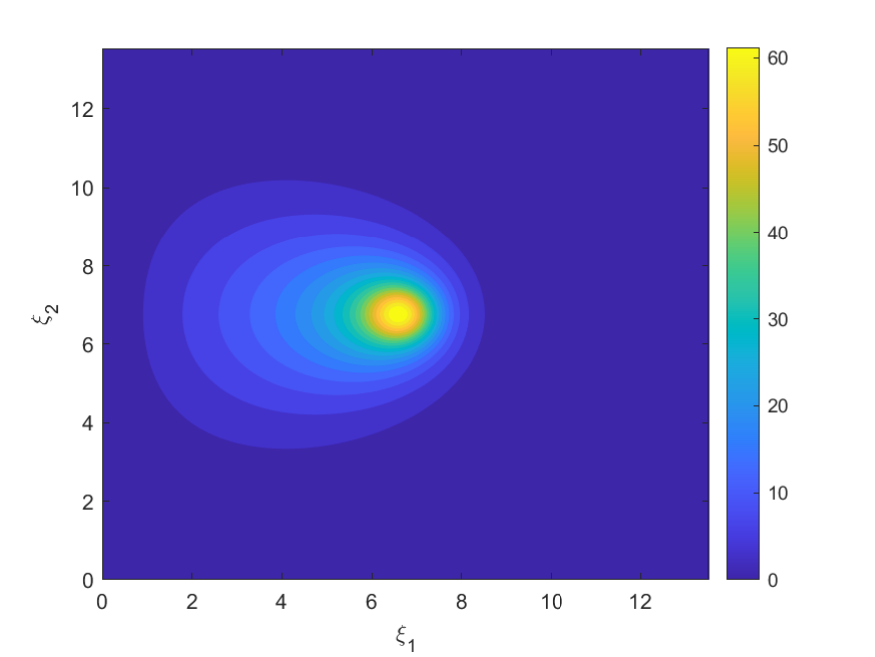}}\quad \quad \quad
	\subfigure[The contour of the temperature in the $\xi_{3}-$direction when $\xi_{2}=W/2$.]
	{\includegraphics[width=0.35\textwidth]{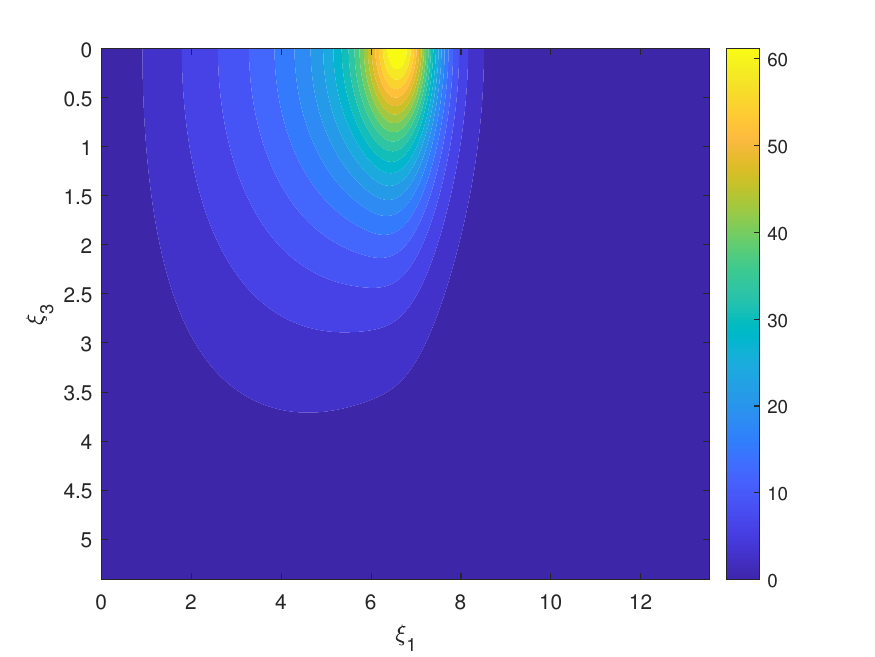}}
	\caption{Temperature distribution of the circular laser beam moving to the center of the 3D medium under the DPL model $(K_{\tau}=1)$.}
	\label{fig:yuan_1-1}
\end{figure}

\begin{figure}[!htbp]
	\centering
	\subfigure[The 3D temperature distribution on the surface of the medium.]
	{\includegraphics[width=0.4\textwidth]{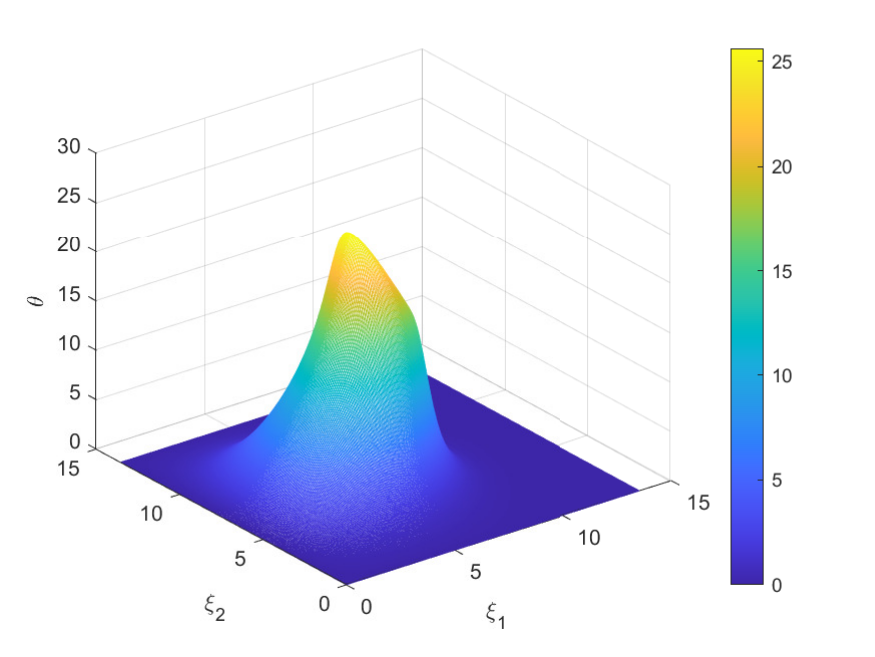}}\\
	\subfigure[The contour of the temperature on the top surface.]
	{\includegraphics[width=0.35\textwidth]{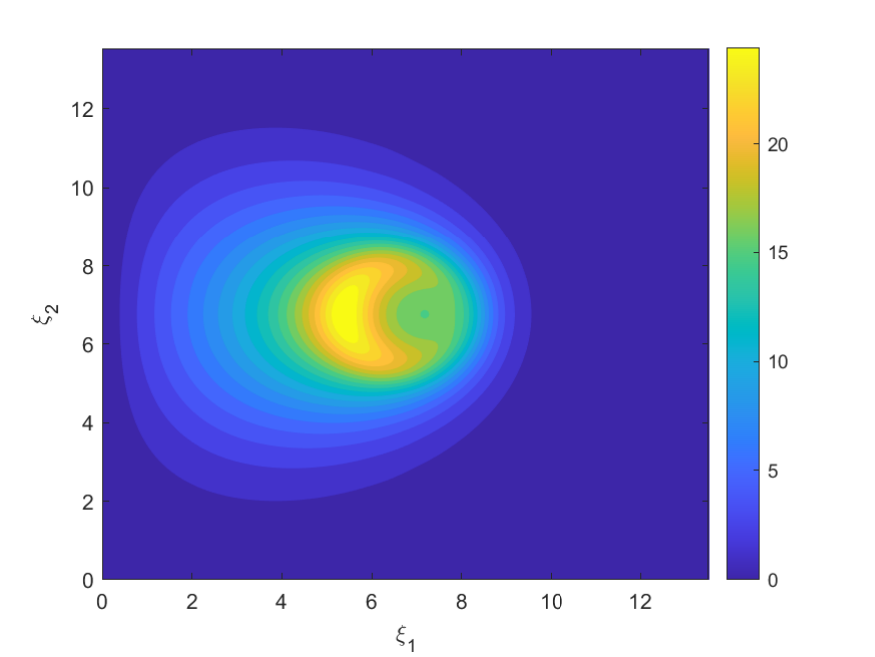}}\quad \quad \quad
	\subfigure[The contour of the temperature in the $\xi_{3}-$direction when $\xi_{2}=W/2$.]
	{\includegraphics[width=0.35\textwidth]{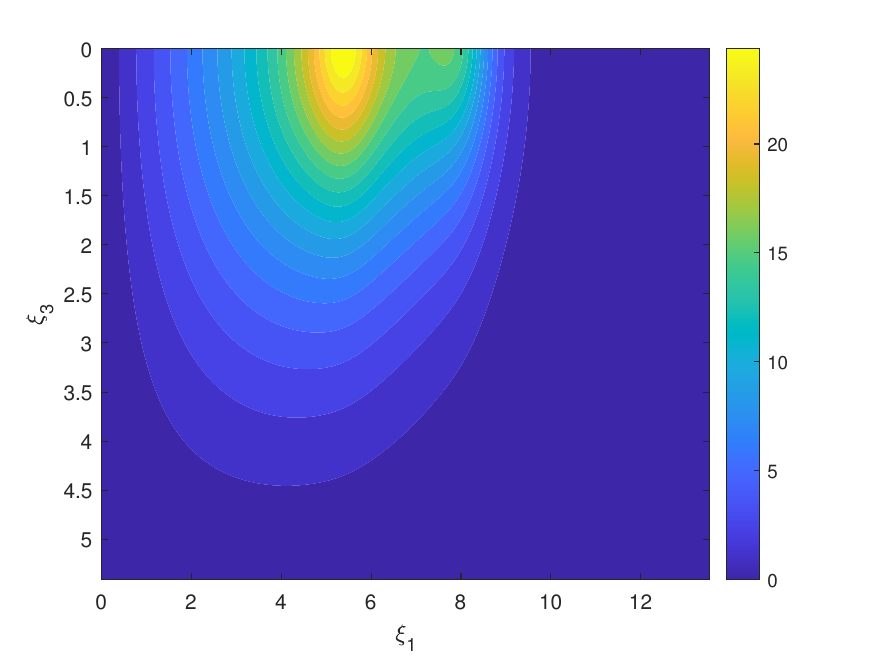}}
	\caption{Temperature distribution of the annular laser beam moving to the center of the 3D medium under the DPL model $(K_{\tau}=1)$.}
	\label{fig:huan_1-1}
\end{figure}

The three-dimensional plot of the temperature distribution on the medium, contours of the top surface and interior of the medium at $K_{\tau}=1$ are presented in \cref{fig:yuan_1-1} for the circular laser beam and in \cref{fig:huan_1-1} for the annular laser beam, respectively. 
When the circular laser beam moving uniformly to the right along the $\xi_{1}$-direction from half of the width of the medium, in \cref{fig:yuan_1-1}, we can observe the temperature in the medium increases with the movement of the laser beam and the temperature peak is located at the place where the laser beam stays. It is also obvious from two contour plots that the heat shows a tendency to diffuse outward in \cref{fig:yuan_1-1}. By the transfer and diffusion of heat from the laser beam moving along the medium, the affected area which means the region is directly heated and subjected to heat conduction gradually increases, and the range subject to heat transfer becomes large with the movement of the laser beam. 
A moving annular laser beam as shown in \cref{fig:huan_1-1} is considered by changing the inner and outer radius. As the laser beam moves, the ring shape is clearly observed on the medium and the heat also diffuses away from the medium. Unlike the circular laser beam, the temperature peak of the annular laser beam is concentrated at the left side of the laser beam, which can be referred to \cite{2022}. As the laser beam moves, the heat is concentrated mainly at the trajectory of the laser beam. The rear region of the annular laser beam on the medium is heated for a longer time than the front, and then the temperature at the rear of the annular is significantly higher than that at the front. When the laser irradiation intensity is the same, the affected area varies with the size of the laser beam. Comparing \cref{fig:yuan_1-1} with \cref{fig:huan_1-1}, it can be seen that the overall temperature of the medium affected by the annular laser beam is lower than that of the medium affected by the circular laser beam. This is because the total energy of the laser beam remains unchanged, and the heat is inversely proportional to the cross-section of the laser beam.

\subsection{Temperature analysis}
In this subsection, the parametric analysis is conducted to reveal the effects of the ratio $K_{\tau}$ of the phase lag parameters, the laser spot size, and the moving laser speed on the temperature distribution of the three-dimensional medium under the action with different moving laser beam.

\subsubsection{Effect of the ratio of the phase lag parameters}
\begin{figure}[!htbp]
	\centering
	\subfigure[Circular laser beam with $K_{\tau}=0.1$.]
	{\includegraphics[width=0.35\textwidth]{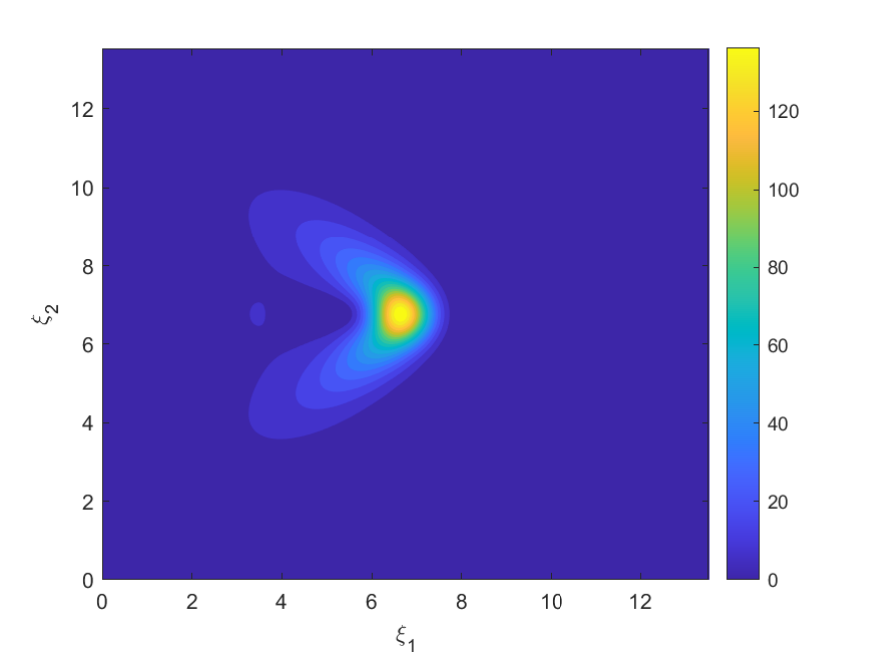}\label{fig:K0.1}} 
	\subfigure[Annular laser beam with $K_{\tau}=0.1$.]
	{\includegraphics[width=0.35\textwidth]{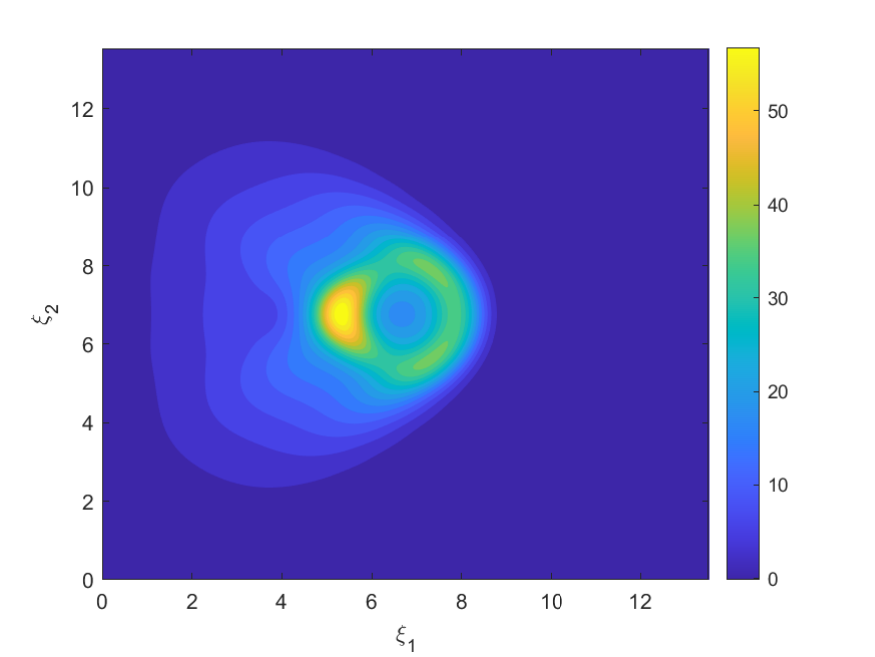}}\\
	\subfigure[Circular laser beam with $K_{\tau}=0.2$.]
	{\includegraphics[width=0.35\textwidth]{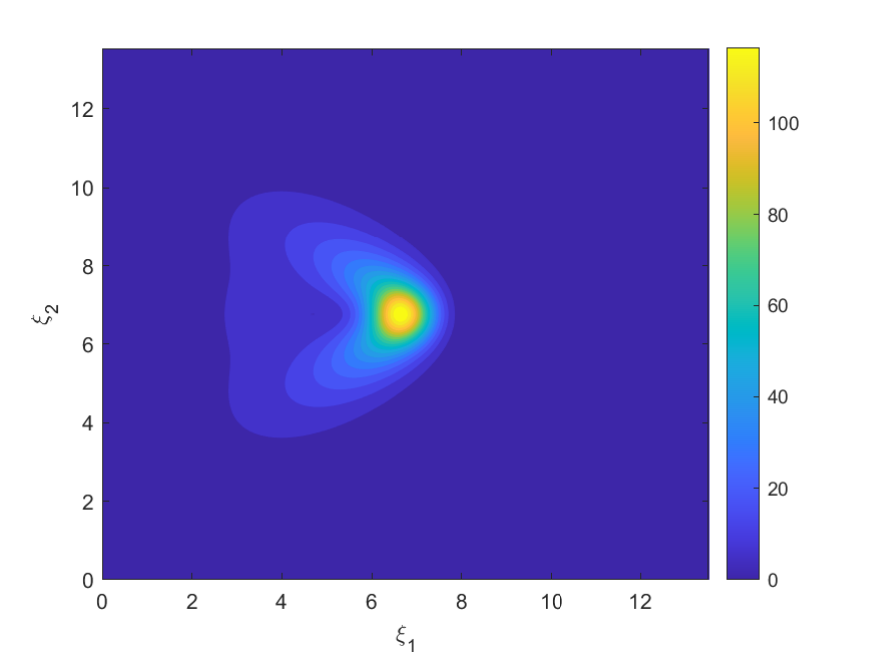}\label{fig:K0.2}} 
	\subfigure[Annular laser beam with $K_{\tau}=0.2$.]
	{\includegraphics[width=0.35\textwidth]{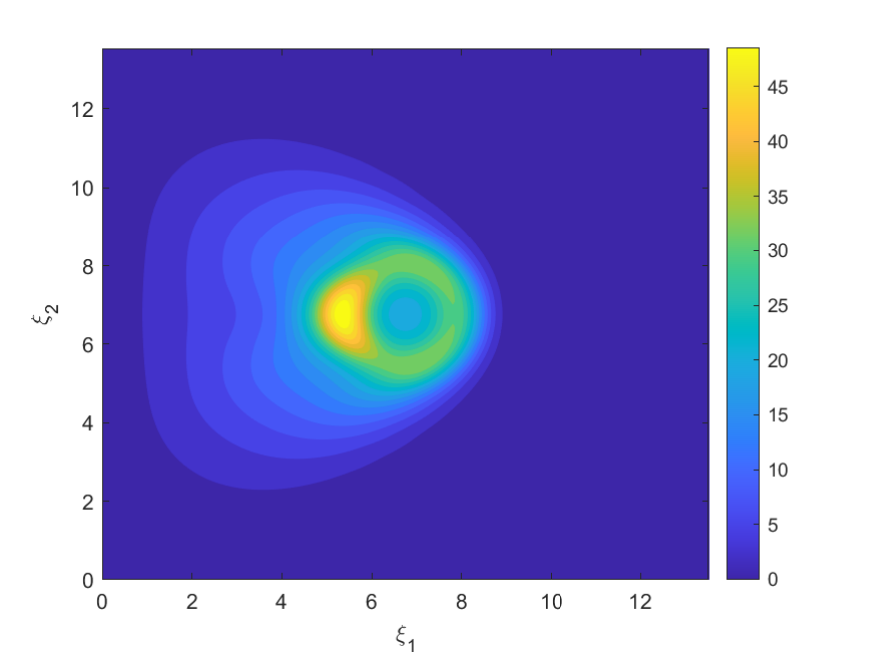}}\\
	\subfigure[Circular laser beam with $K_{\tau}=5$.]
	{\includegraphics[width=0.35\textwidth]{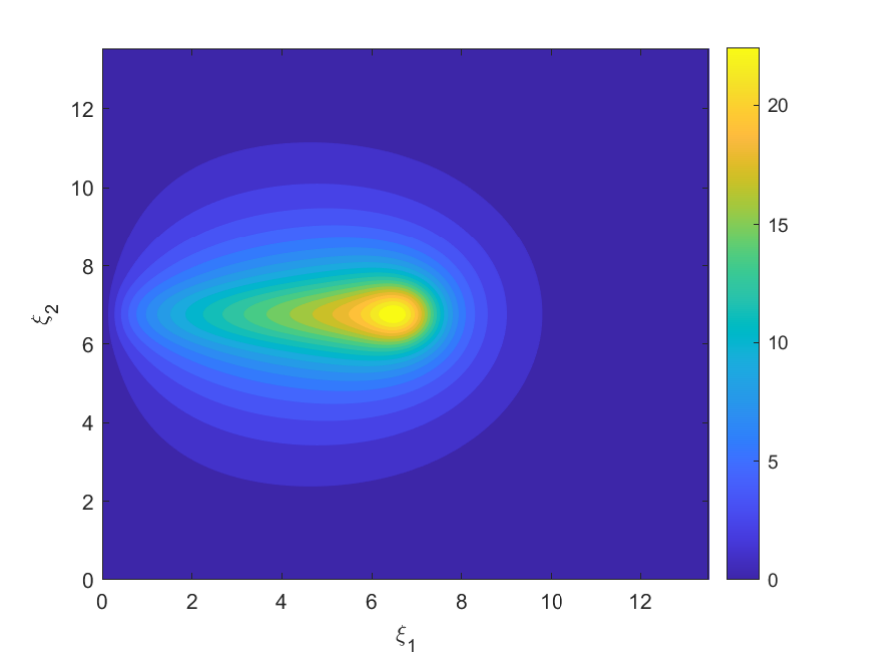}}
	\subfigure[Annular laser beam with $K_{\tau}=5$.]
	{\includegraphics[width=0.35\textwidth]{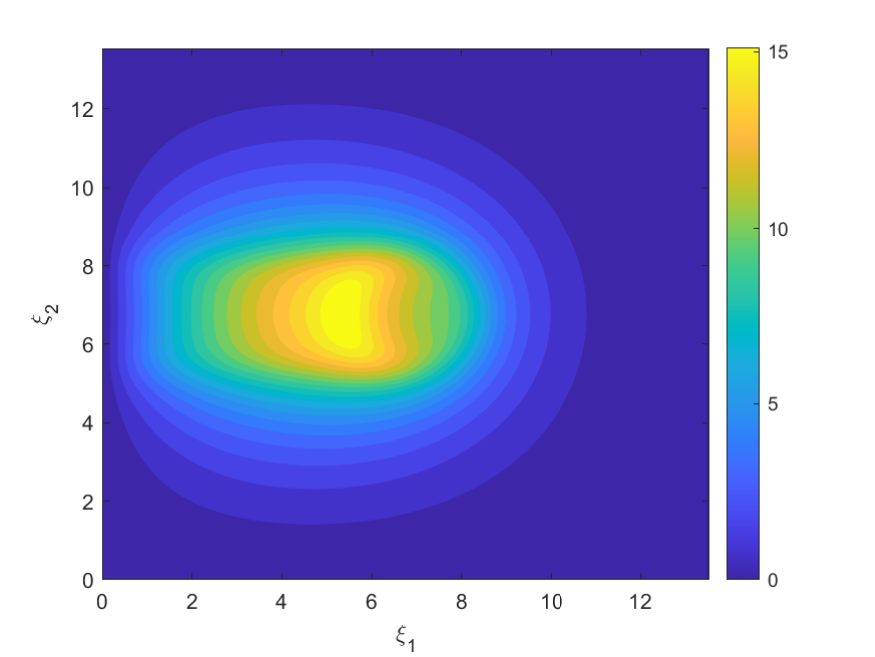}}\\
	\subfigure[Circular laser beam with $K_{\tau}=10$.]
	{\includegraphics[width=0.35\textwidth]{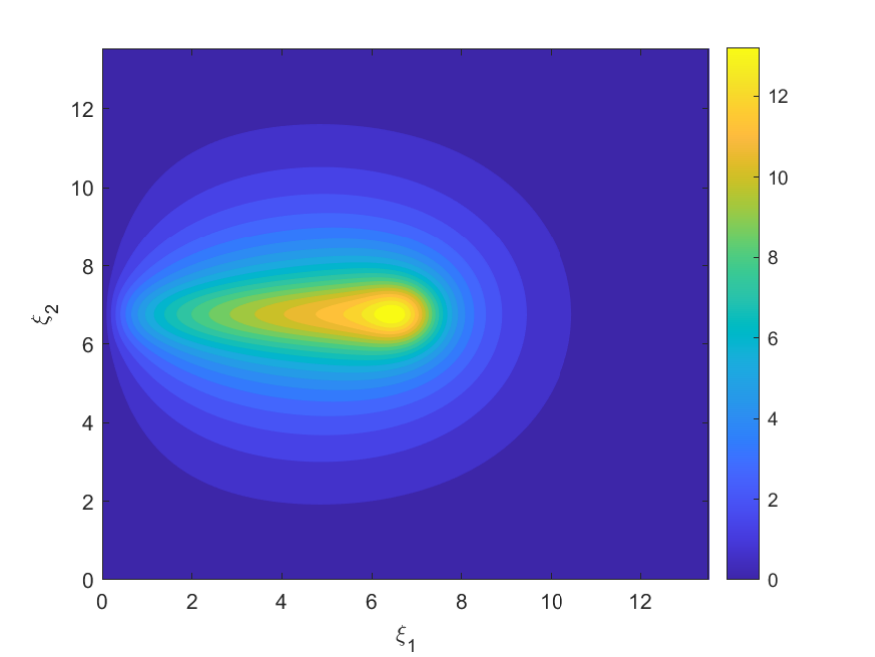}}
	\subfigure[Annular laser beam with $K_{\tau}=10$.]
	{\includegraphics[width=0.35\textwidth]{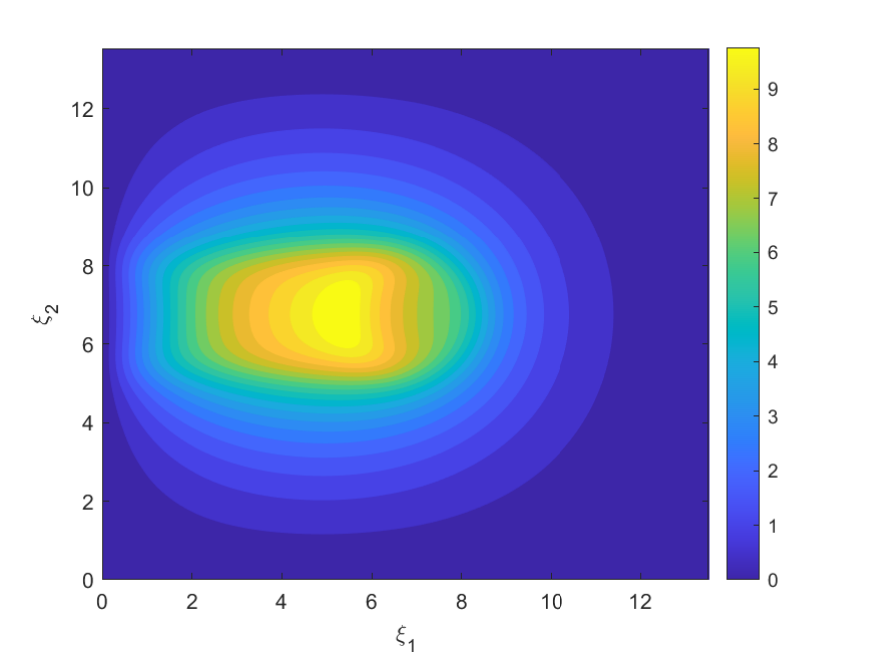}}\\
	\caption{Contours of the temperature distribution on the top surface of the 3D medium for different values of $K_{\tau}$.}
	\label{fig:diff_phase} 
\end{figure}
The variations in temperature distribution over the medium are observed by varying the ratio $K_{\tau}$ on the basis of the DPL model. The general expression for temperature is shown in \cref{eq:series-truncated-sol} and the shape of the laser beam is changed by substituting different parameters of the inner and outer radius as listed in \cref{tab:HS_para}. 
Here we still consider the moving laser beams with circular and annular cross-sections.
The contour plots of the surface of the medium subjected to a circular laser beam with four different ratios of phase lags are presented on the left side of \cref{fig:diff_phase}. As the laser beam moves from left to right along the medium, the temperature peak on the medium appears at the direct heating space. When $K_{\tau}$ is $0.1$ and $0.2$, the overall temperature on the medium does not show a gradual increase on the trajectory which observed in \cref{fig:yuan_1-1}. The obvious phenomenon is that there is a low-temperature zone on the trajectory of the laser beam movement. Such a low-temperature zone makes the original uniform diffusion of heat split up and down, resulting in the temperature decreasing rapidly to the left of the irradiated area. 
While the temperature of the upper left corner and of the lower left corner is gradually declining, a ``butterfly" shape temperature distribution appears. Different from the \cref{fig:K0.1,fig:K0.2}, when $K_{\tau}$ is taken as $5$ and $10$, the heat is concentrated on the trajectory of the laser beam movement and spreads uniformly around, and the maximum temperature on the medium is relatively low.
A similar phenomenon can be observed in the four pictures of the action of the annular laser beam on the right. When $K_{\tau} = 0.1$ and $0.2$, the ``butterfly" shape region on the left also appears. Due to the peculiarity of the annular laser beam, there is no heat in the middle part of the ring, the medium has a low-temperature zone in the middle of the ring. Another low-temperature zone is concentrated in the laser beam trajectory. The highest temperature on the medium subjected to the annular laser beam is still located at the rear of the ring, and the heat spreads around. As $K_{\tau}$ increases, the low-temperature zone gradually disappears. When $K_{\tau}$ takes $5$ and $10$, the heat on the medium is gradually distributed evenly, and the area of the high-temperature region determined by the outer radius of the ring is relatively large. Compared with the left and right figures, the temperature distribution of the medium subjected to the action of the annular laser beam is more complicated.

\begin{figure}[!htbp]
	\centering
	\includegraphics[width=0.4\textwidth]{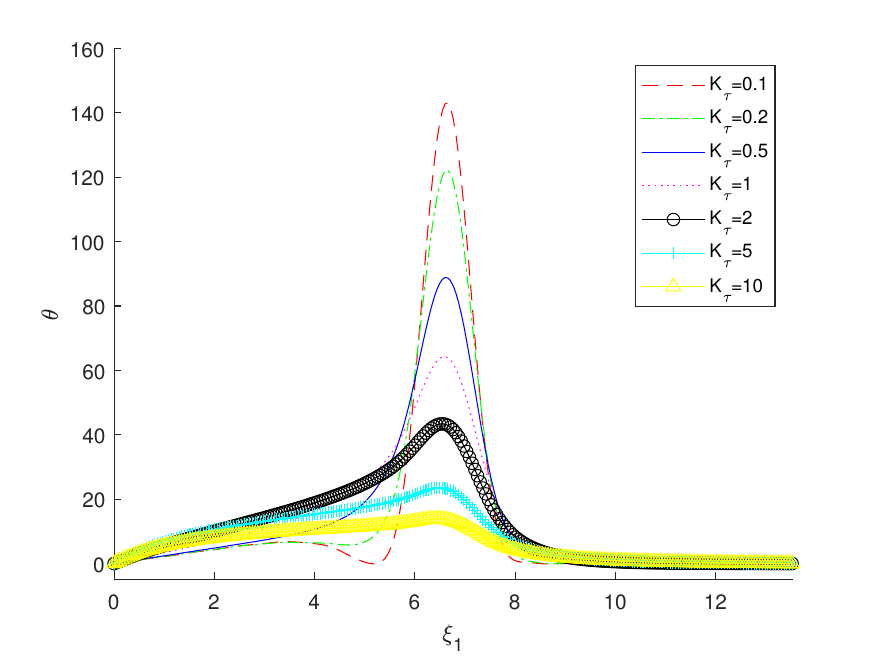}
	\includegraphics[width=0.4\textwidth]{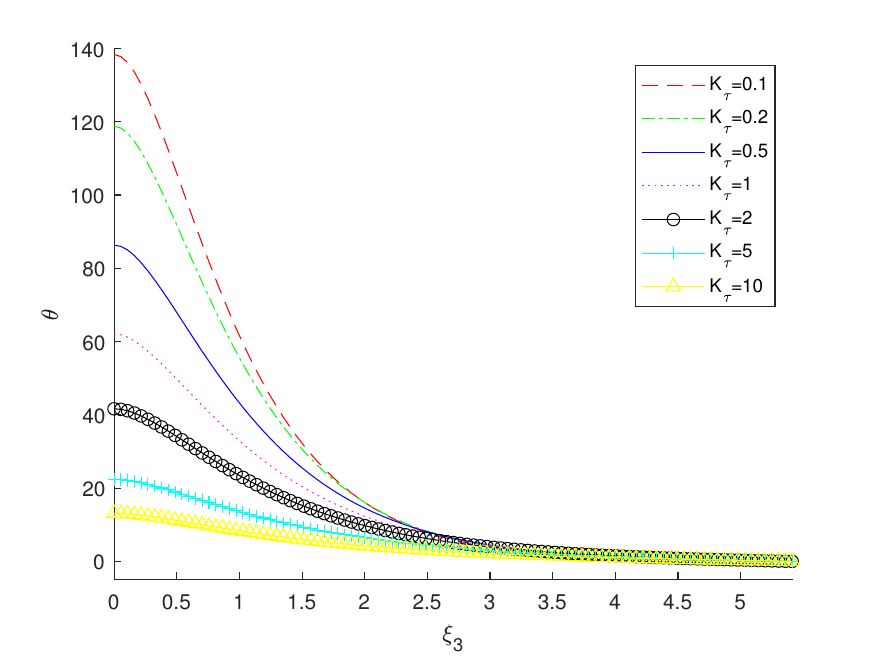}	
	\caption{Top surface temperatures in the $\xi_{1}$-direction(left) and temperatures in the $\xi_{3}$-direction(right) of the medium subjected to a circular laser beam under different $K_{\tau}$.}
	\label{fig:yuan_com}
\end{figure}

\begin{figure}[!htbp]
	\centering
	\includegraphics[width=0.4\textwidth]{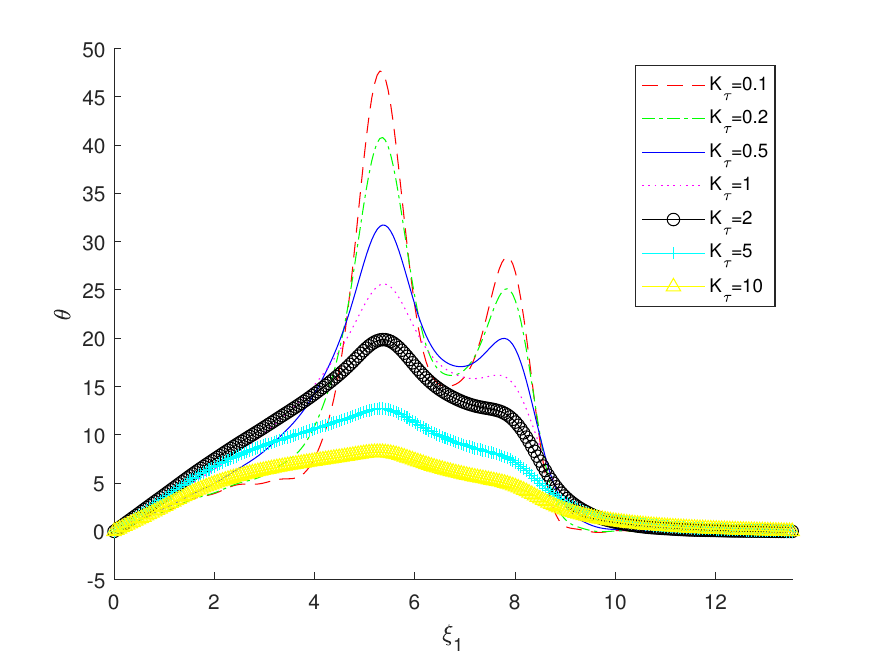}
	\includegraphics[width=0.4\textwidth]{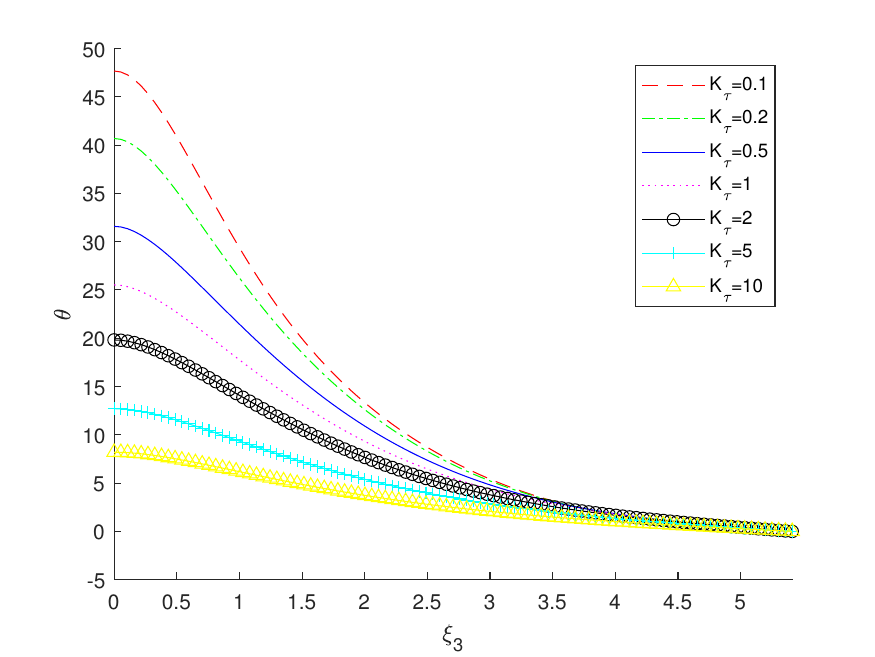}	
	\caption{Top surface temperatures in the $\xi_{1}$-direction(left) and temperatures in the $\xi_{3}$-direction(right) of the medium subjected to an annular laser beam under different $K_{\tau}$.}
	\label{fig:huan_com}
\end{figure}

To further analyze the temperature of the surface and interior of the medium as influenced by the two phase lags, the one-dimensional temperature graphs of the surface and interior can be observed separately in \cref{fig:yuan_com} and \cref{fig:huan_com}. When the cross-section of the laser beam is a circle, the variation of temperature in the $\xi_{1}$-direction for different ratios over half of the width of the medium is demonstrated on the left side of \cref{fig:yuan_com}. From this figure, it can be found that the temperature peak decreases as $K_{\tau}$ increases. It was previously introduced that the DPL model degenerates to a Fourier model when the ratios of the two phase lags are equal. As shown by the red dashed line $K_{\tau} = 0.1$, there is a low-temperature valley in front of the peak, and this temperature valley gradually disappears as $K_{\tau}$ becomes progressively large. The reason for the existence of this phenomenon is that the heat flux phase lag parameter $\tau_{q}$ here is larger than the temperature gradient phase lag $\tau_{T}$.
 A long time lag in the heat flux response rate is caused by a large $\tau_{q}$. Therefore with the  variations of the temperature gradient, the change in heat flux may last for a long time and causes fluctuations in the medium. A recognizable low-temperature region at the rear of the heated region is caused by such a wave-like heat conduction model. When $\tau_{q}$ is large, the heat flux is severely lagged, which reduces the efficiency of heat transfer and results in heat accumulation in the irradiated area. The larger $\tau_{q}$ is, the smaller $K_{\tau}$ is, and the larger the temperature peak in the irradiated area.
If $K_{\tau} \textgreater 1$, it is found from the figure that the temperature distribution over the whole area is more uniform and tends to an equilibrium state as $K_{\tau}$ is larger, which corresponds to \cref{fig:diff_phase}. When $\tau_{T} \textgreater \tau_{q}$, the phase lag of the temperature gradient means that the effect of temperature on the reduction of heat flux is delayed and the thermal equilibrium process is achieved earlier than Fourier's law. The increment of $\tau_{T}$ leads to a faster heat transfer process and an earlier thermal equilibrium. Therefore, the temperature gradient lag promotes the heat transfer process, and the larger the $\tau_{T}$, the lower the temperature peak.
Taking the center of the medium surface as the observation point, we can compare the temperature inside the medium for different values of phase lag ratio from the right side of \cref{fig:yuan_com}. Echoing the plot on the left, the temperature decreases as $K_{\tau}$ increases. However, no matter how the phase lag parameter changes, roughly at the center of the interior, the temperature inside the medium is no longer influenced by the laser beam and gradually tends to 0.

When the laser beam is in the form of a ring, it is able to compare the temperature variation in the $\xi_{1}$-direction for seven different ratios from the left side of \cref{fig:huan_com}. As presented above, the temperature peaks are concentrated at the back of the ring. When $K_{\tau} \textless 1$, the heat flux phase lag $K_{\tau}$ becomes progressively large as the ratio decreases and the two low-temperature regions on the medium become progressively clear. The heat flux rebalances the temperature distribution by transferring heat from the high-temperature region to the low-temperature region. Therefore, the heat flux delays the thermal equilibrium, allowing more heat to accumulate at the laser irradiation area, and then the temperature peak increases. When $K_{\tau} \textgreater 1$, the temperature of the whole medium tends to equilibrate with a large ratio. As the ratio is 10, the two peaks of the ring are not even visible in the left side of \cref{fig:huan_com}. The temperature inside the medium at the point of the highest crest is shown on the right. Similar to \cref{fig:yuan_com}, as the ratio increases, the heat spreads more uniformly and the temperature decreases.

\subsubsection{Effect of laser spot size}
\begin{figure}[!htbp]
	\centering
	\includegraphics[width=0.4\textwidth]{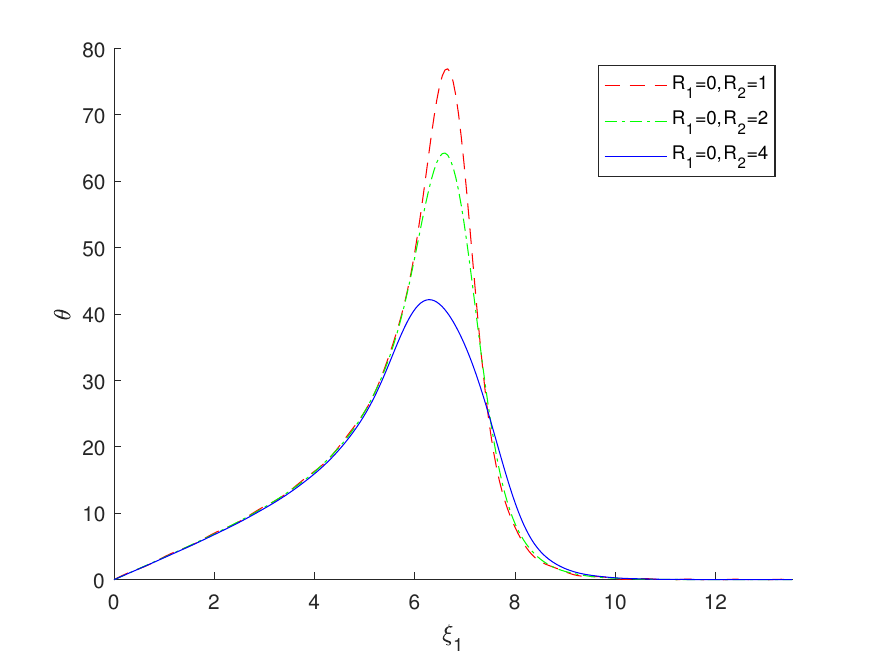}
	\includegraphics[width=0.4\textwidth]{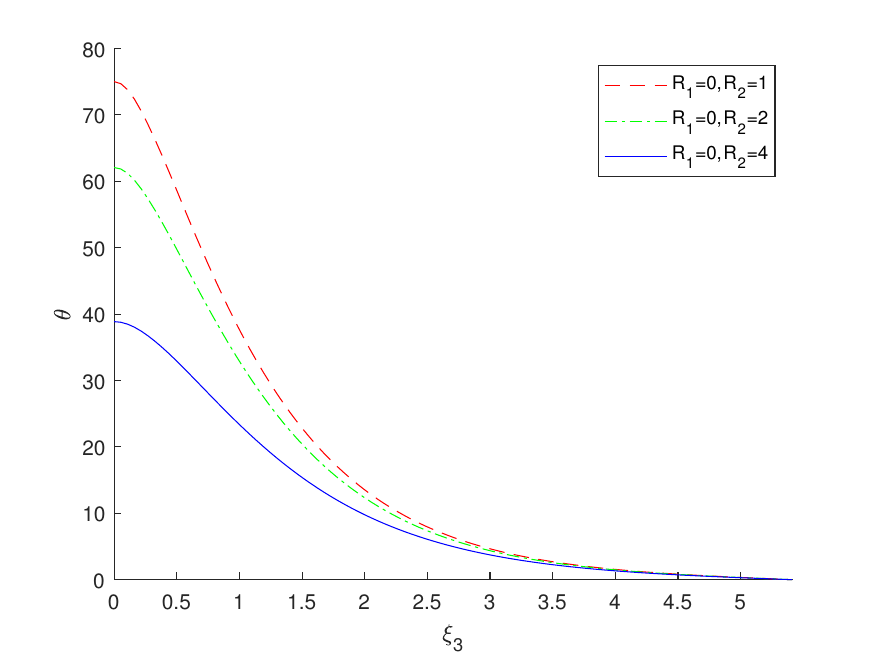}
	\caption{Top surface temperatures in the $\xi_{1}$-direction(left) and temperatures in the $\xi_{3}$-direction(right) of the medium subjected to different sizes of circular laser beams under $K_{\tau}=1$.}
	\label{fig:yuan_com_spot}
\end{figure}

\begin{figure}[!htbp]
	\centering
	\includegraphics[width=0.4\textwidth]{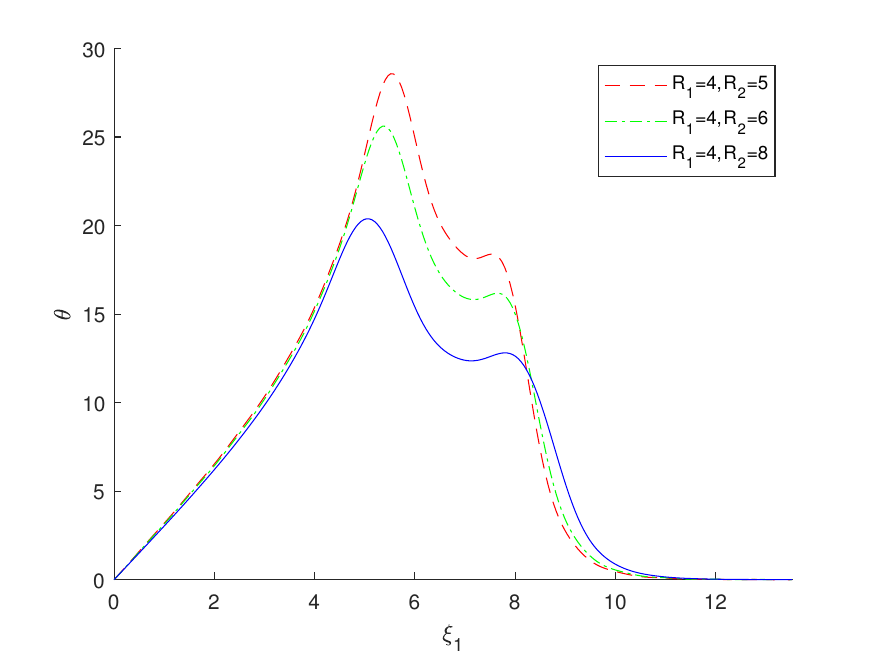}
	\includegraphics[width=0.4\textwidth]{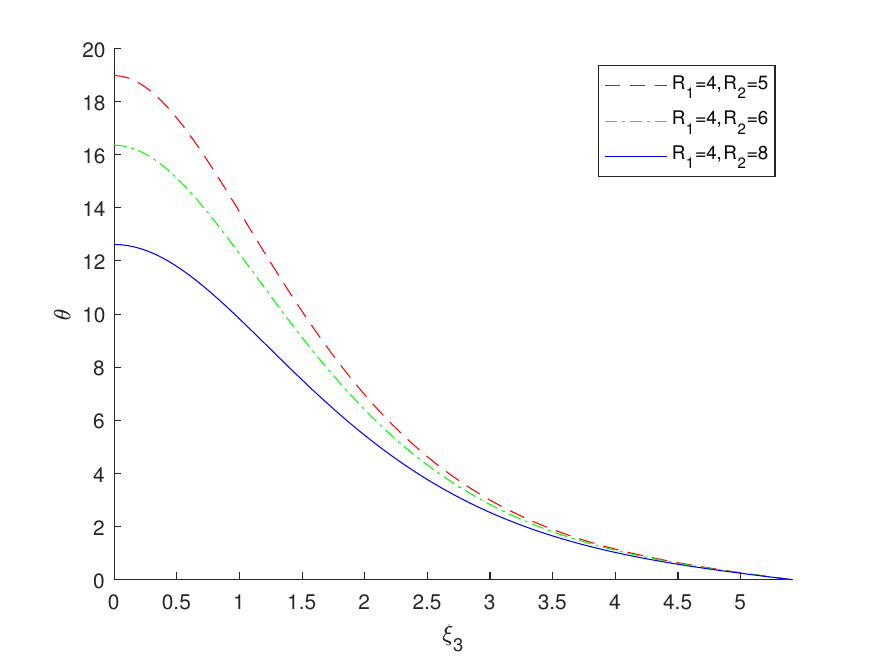}
	\caption{Top surface temperatures in the $\xi_{1}$-direction(left) and temperatures in the $\xi_{3}$-direction(right) of the medium subjected to different sizes of annular laser beams under $K_{\tau}=1$.}
	\label{fig:huan_com_spot}
\end{figure}

The size of the laser spot also has a significant effect on the temperature. The temperature changes on the medium by varying the spot size under the action of two shaped laser beams are shown in \cref{fig:yuan_com_spot,fig:huan_com_spot}, respectively.
 Here the phase lag ratio $K_{\tau}$ is chosen to be 1, and both laser beams move to the center of the medium. It is assumed that the total energy of the laser beam remains unchanged when the size of the laser spot changes during the heating process. Therefore, the energy density is inversely proportional to the area of the laser spot at this time. In both \cref{fig:yuan_com_spot} and \cref{fig:huan_com_spot}, it can be observed that the temperature of the surface and the interior of the medium decreases as the area of the spot becomes larger. This is due to the fact that a large laser spot also expands the affected area on the surface of the medium, promoting the diffusion of heat, and then the temperature peak at the center decreases.

\subsubsection{Effect of moving speed of laser spot}
\begin{figure}[!htbp]
	\centering
	\includegraphics[width=0.4\textwidth]{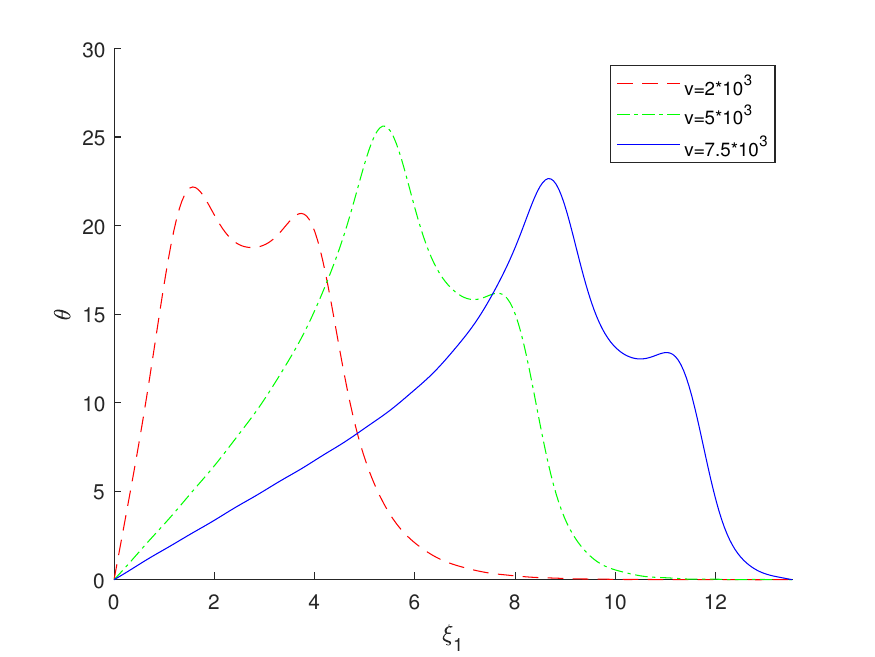}
	\includegraphics[width=0.4\textwidth]{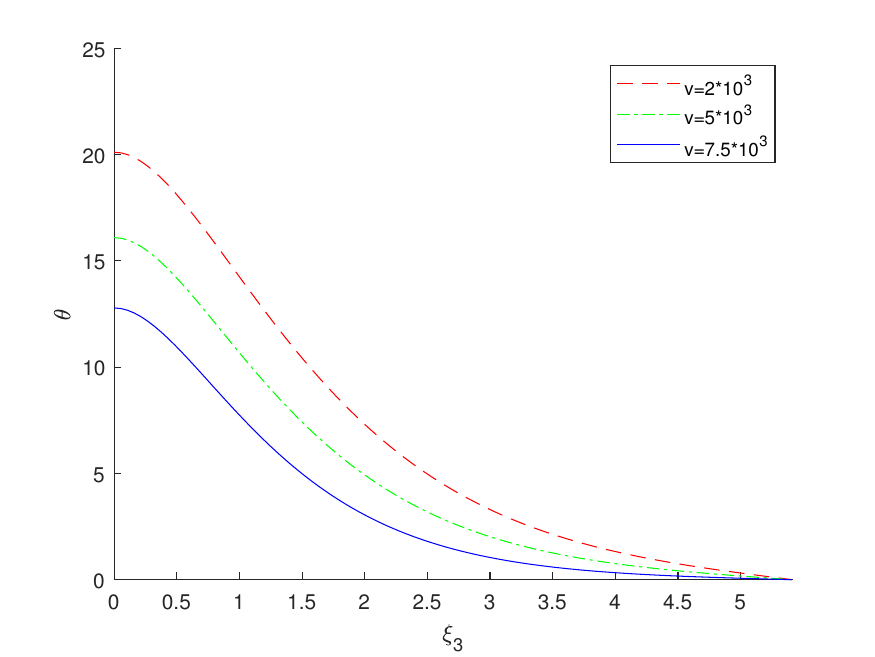}
	\caption{Top surface temperatures in the $\xi_{1}$-direction(left) and temperatures in the $\xi_{3}$-direction(right) of the medium subjected to an annular laser beam with different moving speeds under $K_{\tau}=1$.}
	\label{fig:huan_com_v}
\end{figure}

Finally, we vary the moving speed of the laser beam with equal values of the two phase lag parameters. Examples of circular and annular laser beams are still used. In the area directly heated by the laser beam, the temperature rises sharply and with the passage of time, the heated area moves with the laser beam. As the speed of the laser beam increases, the surface area of the medium heated by the laser beam in the same amount of time expands.
\begin{figure}[H]
	\centering
	\includegraphics[width=0.4\textwidth]{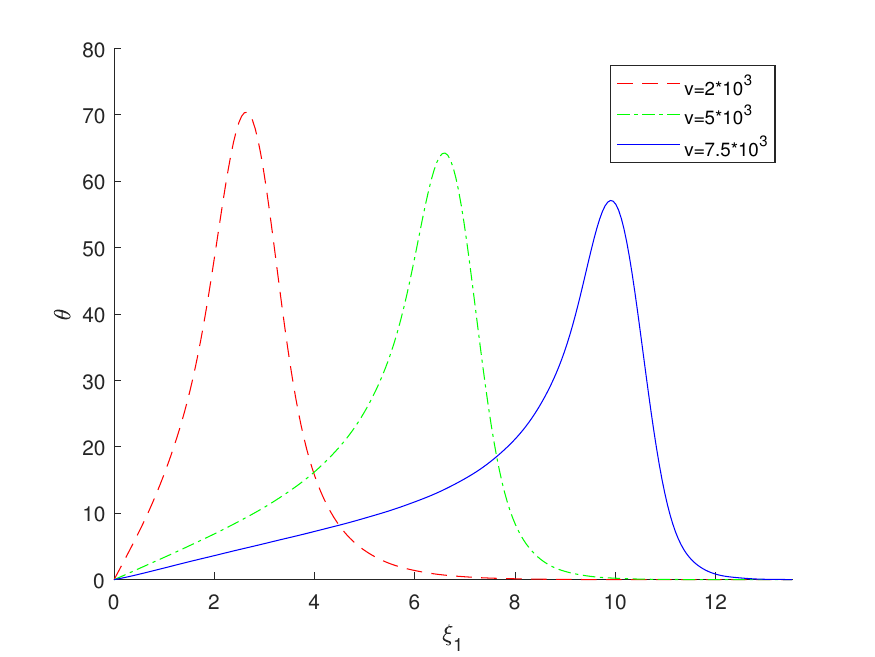}
	\includegraphics[width=0.4\textwidth]{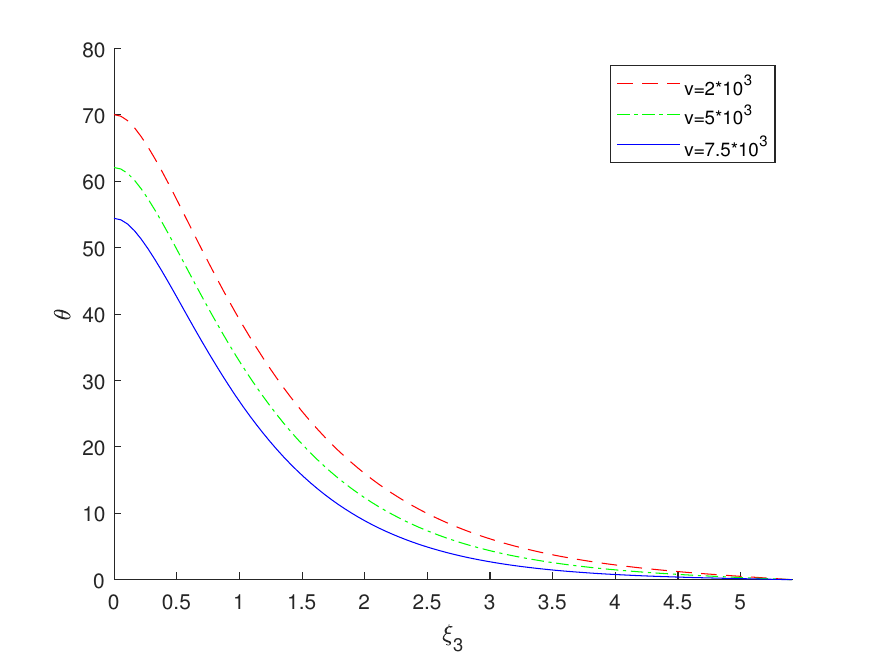}
	\caption{Top surface temperatures in the $\xi_{1}$-direction(left) and temperatures in the $\xi_{3}$-direction(right) of the medium subjected to a circular laser beam with different moving speeds under $K_{\tau}=1$.}
	\label{fig:yuan_com_v}
\end{figure} 
Under the same conditions of laser irradiation intensity, the distance traveled by the laser beam increases with increasing speed, and the heat on the medium also gradually spreads. As can be observed in \cref{fig:yuan_com_v}, the laser beam moves faster, the affected area is larger and the temperature of the surface and the interior of the medium is lower. The same is true for the temperature change inside the 3D medium.
When the cross-section of the laser beam is annular, as seen in \cref{fig:huan_com_v}.  The temperature at the rear of the ring is not as large as when the velocity is high. However, the temperature trend of the front part of the ring irradiated on the medium is the same as that of the circular laser beam. The reason for this phenomenon is that under the same conditions, the affected area expands with increasing speed of laser beam movement. Due to the special shape of the annular laser beam, a large amount of heat is concentrated in the rear half of the ring. The heat is relatively uniform as the speed is slow. When the speed increases, the temperature of the rear part of the ring rises rapidly. While when the speed is relatively faster, the temperature at the rear of the ring decreases because more medium is acted upon and most of the heat spreads around. Observing the front part of the ring, it is found that the trend of temperature change is the same as that of the circular laser beam. As shown on the right side of \cref{fig:huan_com_v}, the temperature inside the medium is lower as the moving speed is faster.

\section{Conclusion}
\label{sec:conclusion}
The temperature distribution in a three-dimensional medium subjected to a moving circular or annular laser beam based on the DPL model is investigated. Different from the previous 2D model, we can observe the temperature distribution on the surface and inside of the medium from the 3D model, which can well simulate the medium heated by the heat source. Green's function method is used to solve for a medium subjected to a single point heat source. Then, according to the different cross-sections of the laser beam, the superposition method is used to obtain the analytical solution of the temperature distribution. After verifying the analytical solution of the DPL model, the temperatures on the medium under Fourier's law and non-Fourier's law are compared from the modeling point of view. Heat transfer analyses are then performed for circular and annular laser beams to investigate the temperature distribution on the surface and inside of the 3D medium. Finally, the effects of phase lag parameter, laser spot size, and laser moving speed on temperature variation are revealed. The main conclusions are as follows:

(1). The value of the phase lag parameter has a large impact on the temperature distribution of the medium. When $K_{\tau} \textless 1$, which means $\tau_{q} \textgreater \tau_{T}$, the heat flux phase lag affects the heat transfer, making the temperature distribution on the whole medium unbalanced. And the heat is concentrated in the heated area, the larger $\tau_{q}$, the higher the temperature at the heated place. As $K_{\tau} = 1$, the temperature distribution under the DPL model is similar to the heat distribution described by Fourier's law. When $K_{\tau} \textgreater 1$, the heat flux precedes the temperature gradient, leading to faster heat transfer and earlier thermal equilibrium, and a more uniform temperature distribution over the medium.

(2). Assuming the same total energy power of the laser, the energy density is inversely proportional to the area of the laser spot. When the laser spot is larger, the temperature peak on the medium is lower instead.

(3). The speed of movement of the laser beam also plays an important role in the temperature distribution of the medium. The faster the laser beam moves, the larger the area of the medium affected by the laser, the more heat spreads outward, and then the temperature peak is reduced.

The above results are useful for analyzing the temperature distribution on the surface and inside of the medium subjected to the action of the moving heat source, which expands the application of the moving heat source in the field of heat conduction.

\section*{Acknowledgements}
This work was partially supported by the National Natural Science
Foundation of China, No. 12171240.
The work of the first author was also partially supported by
Postgraduate Research \& Practice Innovation Program of NUAA,
No. xcxjh20220803.
	
	\appendix
	\section{Calculation of integral in \cref{eq:series-truncated-sol}}
	\label{appendix}
	The integral in \cref{eq:series-truncated-sol} can be written as
	\begin{equation}
		\label{eq:integral_xy}
		\int_{R_1}^{R_2}\int_{0}^{2\pi}\sin (\alpha_{m}(\xi_{1}-R_0\cos \theta))\sin(\beta_{n}(\xi_{2}-R_0\sin \theta))R_0\dd R_0 \dd \theta ,
	\end{equation}
	where
	\begin{equation}
		\label{eq:sin_sin}
		\begin{aligned}
			&\sin (\alpha_{m}(\xi_1 -R_0\cos \theta ))\sin(\beta_{n}(\xi_2-R_0\sin\theta))=\\
			&\sin (\alpha_{m}\xi_1)\sin(\beta_{n}\xi_2) \cos(\alpha_{m}R_0\cos\theta)\cos(\beta_{n}R_0\sin\theta)+\cos(\alpha_{m}\xi_1)\cos(\beta_{n}\xi_2)\sin(\alpha_{m}R_0\cos\theta)\sin(\beta_{n}R_0\sin\theta)\\
			&-\cos(\alpha_{m}\xi_1)\sin(\beta_{n}\xi_2)\sin(\alpha_{m}R_0\cos\theta)\cos(\beta_{n}R_0\sin\theta)-\sin (\alpha_{m}\xi_1)\cos(\beta_{n}\xi_2) \cos(\alpha_{m}R_0\cos\theta)\sin(\beta_{n}R_0\sin\theta).
		\end{aligned}
	\end{equation}
	Using the following relationships
	\begin{subequations}
		\begin{align}
			&\cos (r \sin  \theta )=J_{0}(r)+ 2 \sum_{i=1}^{\infty}  J_{2 i}(r) \cos 2 i  \theta,\label{eq:cs} \\
			&\sin(r\sin  \theta  )=2\sum_{i=0}^{\infty} J_{2i+1}(r)\sin (2i+1) \theta, \label{eq:ss} \\
			&\cos(r\cos  \theta )=J_0(r)+2\sum_{i=1}^{\infty}(-1)^i J_{2i}(r)\cos 2 i  \theta, \label{eq:cc}\\
			&\sin(r\cos  \theta )=2\sum_{i=0}^{\infty} (-1)^i J_{2i+1}(r)\cos(2i+1) \theta, \label{eq:sc}
		\end{align}
	\end{subequations}
	we can obtain
	\begin{subequations}
		\begin{align}
			&\cos(A\cos \theta )\cos(B\sin \theta )=4 \sum_{i=0}^{\infty} \sum_{j=0}^{\infty}(-1)^{i} \chi_{i} \chi_{j} J_{2 i}(A) J_{2 j}(B) \cos 2 i   \theta  \cos 2 j  \theta,\label{eq:cccs} \\
			&\cos(A\cos \theta )\sin(B\sin \theta )=4 \sum_{i=0}^{\infty} \sum_{j=0}^{\infty}(-1)^{i} \chi_{i} J_{2 i}(A) J_{2 j+1}(B) \cos 2 i  \theta  \sin (2 j+1)   \theta,\label{eq:ccss} \\
			&\sin(A\cos \theta )\cos(B\sin \theta )=4 \sum_{i=0}^{\infty} \sum_{j=0}^{\infty}(-1)^{i} \chi_{j} J_{2 i+1}(A) J_{2 j}(B) \cos (2 i+1)   \theta  \cos 2 j  \theta,\label{eq:sccs} \\
			&\sin(A\cos \theta )\sin(B\sin \theta )=4 \sum_{i=0}^{\infty} \sum_{j=0}^{\infty}(-1)^{i} J_{2 i+1}(A) J_{2 j+1}(B) \cos (2 i+1)  \theta  \sin (2 j+1)   \theta,\label{eq:scss} 
		\end{align}
	\end{subequations}
	where $\chi_{0}=0.5$ and $\chi_{i}=1$ with $i=1,2,...$. 
	
	According to the orthogonality of trigonometric functions, the integral value of \cref{eq:ccss,eq:sccs,eq:scss} for angle is $0$, and the integral values of \cref{eq:cccs} can be simplified as
	\begin{equation}
		\label{eq:cccs_angle}
		\int_{0}^{2\pi}\cos(A\cos \theta )\cos(B\sin \theta )\dd \theta =2\pi J_0(A)J_0(B) +4\pi\sum_{i=1}^{\infty}(-1)^i J_{2i}(A)J_{2i}(B).
	\end{equation}
	Taking the integral \cref{eq:cccs_angle} over the angle into the double integral \cref{eq:integral_xy}, we get 
	\begin{equation}
		\label{eq:appendix_final}
		\begin{aligned}
			&\int_{R_1}^{R_2}\int_{0}^{2\pi} \sin (\alpha_{m}(\xi_{1}-R_0\cos \theta))\sin(\beta_{n}(\xi_{2}-R_0\sin \theta)) R_0\dd R_0 \dd \theta \\
			=&\int_{R_1}^{R_2}\int_{0}^{2\pi}\sin (\alpha_{m}\xi_1)\sin(\beta_{n}\xi_2) \cos(\alpha_{m}R_0\cos\theta)\cos(\beta_{n}R_0\sin\theta)\dd R_0 \dd \theta \\
			=&\sin (\alpha_{m}\xi_1)\sin(\beta_{n}\xi_2) \int_{R_1}^{R_2}\left(2\pi R_0J_0(\alpha_{m}R_0)J_0(\beta_{n}R_0)+4\pi\sum_{i=1}^{\infty}
			(-1)^iR_0J_{2i}(\alpha_{m}R_0)J_{2i}(\beta_{n}R_0)\right)\dd R_0,\\
		\end{aligned}
	\end{equation}
	where 
	\begin{equation}
		\int_{R_1}^{R_2}R_0J_0(\alpha_{m}R_0)J_0(\beta_{n}R_0)\dd R_0=
		\begin{cases}
			\frac{1}{2}(-R_1^2(J_0(\alpha_{m}R_1)^2+J_1(\alpha_{m}R_1)^2)+R_2^2(J_0(\alpha_{m}R_2)^2+J_1(\alpha_{m}R_2)^2)), &   \alpha_{m} = \beta_{n}, \\	
			\\
			\begin{aligned}
				&\frac{1}{\alpha_{m}^2-\beta_{n}^2}
				(-\alpha_{m}R_1J_0(\beta_{n}R_1)J_1(\alpha_{m}R_1)+\beta_{n}R_1J_0(\alpha_{m}R_1)J_1(\beta_{n}R_1)\\
				&+\alpha_{m}R_2J_0(\beta_{n}R_2)J_1(\alpha_{m}R_2)-\beta_{n}R_2J_0(\alpha_{m}R_2)J_1(\beta_{n}R_2)),
			\end{aligned} & \alpha_{m} \neq \beta_{n},
		\end{cases}
	\end{equation}
	and for $i \geq 1$,
	\begin{equation}
		\int_{R_1}^{R_2}R_0J_{2i}(\alpha_{m}R_0)J_{2i}(\beta_{n}R_0)\dd R_0=
		\begin{cases}
			\begin{aligned}
				&\frac{1}{2\alpha_{m}}(-\alpha_{m}R_1^2J_{2i}^2(\alpha_{m}R_1)+\alpha_{m}R_2^2J_{2i}^2(\alpha_{m}R_2)+\alpha_{m}R_2^2J_{2i-1}^2(\alpha_{m}R_2)\\&-\alpha_{m}R_1^2J_{2i-1}^2(\alpha_{m}R_1)+4i(R_1J_{2i}(\alpha_{m}R_1)J_{2i-1}(\alpha_{m}R_1)\\&-R_2J_{2i}(\alpha_{m}R_2)J_{2i-1}(\alpha_{m}R_2))),
			\end{aligned} & \alpha_{m} = \beta_{n}, \\	
			\\
			\begin{aligned}
				&\frac{1}{\alpha_{m}^2-\beta_{n}^2}
				(\alpha_{m}R_1J_{2i}(\beta_{n}R_1)J_{2i-1}(\alpha_{m}R_1)-\beta_{n}R_1J_{2i}(\alpha_{m}R_1)J_{2i-1}(\beta_{n}R_1)\\&-\alpha_{m}R_2J_{2i}(\beta_{n}R_2)J_{2i-1}(\alpha_{m}R_2)+\beta_{n}R_2J_{2i}(\alpha_{m}R_2)J_{2i-1}(\beta_{n}R_2)),
			\end{aligned} &  \alpha_{m} \neq \beta_{n}.
		\end{cases} 
	\end{equation}
	Therefore, the integral \cref{eq:integral_xy} can be approximately equal to series \cref{eq:appendix_final}.
	
	\bibliographystyle{elsarticle-num} 
	\bibliography{references.bib}
\end{document}